\documentclass[aps,pra,twocolumn,10pt,letterpaper,superscriptaddress]{revtex4-1}

\usepackage{xcolor, graphicx}
\usepackage{amsmath, amssymb}
\usepackage{ulem}
\usepackage{enumerate}
\usepackage[colorlinks=true,urlcolor=blue,citecolor=blue,linkcolor=blue]{hyperref}

\begin{document}

\title{Identification of nonclassical properties of light with multiplexing layouts}

\author{J.~Sperling}\email{jan.sperling@physics.ox.ac.uk}
\affiliation{Clarendon Laboratory, University of Oxford, Parks Road, Oxford OX1 3PU, United Kingdom}

\author{A.~Eckstein}
\affiliation{Clarendon Laboratory, University of Oxford, Parks Road, Oxford OX1 3PU, United Kingdom}

\author{W.~R.~Clements}
\affiliation{Clarendon Laboratory, University of Oxford, Parks Road, Oxford OX1 3PU, United Kingdom}

\author{M.~Moore}
\affiliation{Clarendon Laboratory, University of Oxford, Parks Road, Oxford OX1 3PU, United Kingdom}

\author{J.~J.~Renema}
\affiliation{Clarendon Laboratory, University of Oxford, Parks Road, Oxford OX1 3PU, United Kingdom}

\author{W.~S.~Kolthammer}
\affiliation{Clarendon Laboratory, University of Oxford, Parks Road, Oxford OX1 3PU, United Kingdom}

\author{S.~W.~Nam}
\affiliation{National Institute of Standards and Technology, 325 Broadway, Boulder, CO 80305, USA}

\author{A.~Lita}
\affiliation{National Institute of Standards and Technology, 325 Broadway, Boulder, CO 80305, USA}

\author{T.~Gerrits}
\affiliation{National Institute of Standards and Technology, 325 Broadway, Boulder, CO 80305, USA}

\author{I.~A.~Walmsley}
\affiliation{Clarendon Laboratory, University of Oxford, Parks Road, Oxford OX1 3PU, United Kingdom}

\author{G.~S.~Agarwal}
\affiliation{Texas A\&M University, College Station, Texas 77845, USA}

\author{W.~Vogel}
\affiliation{Institut f\"ur Physik, Universit\"at Rostock, Albert-Einstein-Stra\ss{}e 23, D-18059 Rostock, Germany}

\date{\today}

\begin{abstract}
	In Ref. \cite{Setal2017}, we introduced and applied a detector-independent method to uncover nonclassicality.
	Here, we extend those techniques and give more details on the performed analysis.
	We derive a general theory of the positive-operator-valued measure that describes multiplexing layouts with arbitrary detectors.
	From the resulting quantum version of a multinomial statistics, we infer nonclassicality probes based on a matrix of normally ordered moments.
	We discuss these criteria and apply the theory to our data which are measured with superconducting transition-edge sensors.
	Our experiment produces heralded multi-photon states from a parametric down-conversion light source.
	We show that the known notions of sub-Poisson and sub-binomial light can be deduced from our general approach, and we establish the concept of sub-multinomial light, which is shown to outperform the former two concepts of nonclassicality for our data.
\end{abstract}

\maketitle

\section{Introduction}
	The bare existence of photons highlights the particle nature of electromagnetic waves in quantum optics \cite{E05}.
	Therefore, the generation and detection of photon states are crucial for a comprehensive understanding of fundamental concepts in quantum physics; see Refs. \cite{BC10,CDKMS14} for recent reviews on single photons.
	Beyond this scientific motivation, the study of nonclassical radiation fields is also of practical importance.
	For instance, quantum communication protocols rely on the generation and detection of photons \cite{GT07,S09}.
	Yet, unwanted attenuation effects---which are always present in realistic scenarios---result in a decrease of the nonclassicality of a produced light field.
	Conversely, an inappropriate detector model can introduce fake nonclassicality even to a classical radiation field \cite{SV11,GLLSSMK11,SVA12a}.
	For this reason, we seek robust and detector-independent certifiers of nonclassicality \cite{Setal2017}.

	The basic definition of nonclassicality is that a quantum state of light cannot be described in terms of classical statistical optics.
	A convenient way to represent general states is given in terms of the Glauber-Sudarshan $P$ function \cite{S63,G63}.
	Whenever this distribution cannot be interpreted in terms of classical probability theory, the thereby represented state is a nonclassical one \cite{TG86,M86}.
	A number of nonclassicality tests have been proposed; see Ref. \cite{MBWLN10} for an overview.
	Most of them are formulated in terms of matrices of normally ordered moments of physical observables; see, e.g., Ref. \cite{SRV05}.
	For example, the concept of nonclassical sub-Poisson light \cite{M79} can be written and even generalized in terms of matrices of higher-order photon-number correlations \cite{AT92}.
	Other matrix-based nonclassicality tests employ the Fourier or Laplace transform of the Glauber-Sudarshan $P$ function \cite{RV02,SVA16}.

	In order to apply such nonclassicality probes, one has to measure the light field under study with a photodetector \cite{S07,H09}.
	The photon statistics of the measured state can be inferred if the used detector has been properly characterized.
	This can be done by a detector tomography \cite{LS99,AMP04,LKKFSL08,LFCPSREPW09,ZDCJEPW12}---i.e., measuring a comparably large number of well-defined probe states to construct a detection model.
	Alternatively, one can perform a detector calibration \cite{BCDGMMPPP12,PHMH12,BKSSV17}---i.e., the estimation of parameters of an existing detection model with some reference measurements.
	Of particular interest are photon-number-resolving detectors of which superconducting transition-edge sensors (TESs) are a successful example \cite{LMN08,Getal11,BCDGLMPRTP12,RFZMGDFE12,ZCDPLJSPW12}.
	Independent of the particular realization, photon-number-resolving devices allow for the implementation of quantum tasks, such as state reconstruction \cite{LCGS10,BGGMPTPOP11}, imaging \cite{LBFD08,CWB14}, random number generation \cite{ATDYRS15}, and the characterization of sources of nonclassical light \cite{WDSBY04,HPHP05,FL13,APHAB16}---even in the presence of strong imperfections \cite{TKE15}.
	Moreover, higher-order \cite{ALCS10,AOB12,SBVHBAS15}, spatial \cite{BDFL08,MMDL12,CTFLMA16}, and conditional \cite{SBDBJDVW16} quantum correlations have been studied.

	So far, we did not distinguish between the detection scheme and the actual detectors.
	That is, one has to discern the optical manipulation of a signal field and its interaction with a sensor which yields a measurement outcome.
	Properly designed detection layouts of such a kind render it possible to infer or use properties of quantum light without having a photon-number-resolution capability \cite{ZABGGBRP05,PADLA10,KV16} or they do not require a particular detector model \cite{CKS14,AGSB16}.
	For instance, multiplexing layouts with a number of detectors that can only discern between the presence (``on'') or absence (``off'') of absorbed photons can be combined into a photon-number-resolving detection device \cite{PTKJ96,KB01,ASSBW03,FJPF03,RHHPH03,CDSM07,SPDBCM07}.
	Such types of schemes use an optical network to split an incident light field into a number of spatial or temporal modes of equal intensities which are subsequently measured with on/off detectors.
	The measured statistics is shown to resemble a binomial distribution \cite{SVA12a} rather than a Poisson statistics, which is obtained for photoelectric detection models \cite{KK64}; see also Refs. \cite{I14,PZA16} in this context.
	For such detectors, the positive-operator-valued measure (POVM), which fully describes the detection layout, has been formulated \cite{SVA12a,MSB16}.
	Recently, the combination of a multiplexing scheme with multiple TESs has been used to significantly increase the maximal number of detectable photons \cite{HBLNGS15}.

	Based on the binomial character of the statistics of multiplexing layouts with on/off detectors, the notion of sub-binomial light has been introduced \cite{SVA12} and experimentally demonstrated \cite{BDJDBW13}.
	It replaces the concept of sub-Poisson light \cite{M79}, which applies to photoelectric counting models \cite{KK64}, for multiplexing arrangements using on/off detectors.
	Nonclassical light can be similarly inferred from multiplexing devices with non-identical splitting ratios \cite{LFPR16}.
	In addition, the on-chip realization of optical networks \cite{MGHPGSWS15} can be used to produce integrated detectors to verify sub-binomial light \cite{HSPGHNVS16}.

	In this paper, we derive the quantum-optical click-counting theory for multiplexing layouts which employ arbitrary detectors.
	Therefore, we formulate nonclassicality tests in terms of normally ordered moments, which are independent of the detector response.
	This method is then applied to our experiment which produces heralded multi-photon states.
	Our results are discussed in relation with other notions of nonclassical photon correlations.

	In Ref. \cite{Setal2017}, we study the same topic as we do in this paper from a classical perspective.
	There, the treatment of the detector-independent verification of quantum light is performed solely in terms of classical statistical optics.
	Here, however, we use a complementary quantum-optical perspective on this topic.
	Beyond that, we also consider higher-order moments of the statistics, present additional features of our measurements, and compare our results with previously known nonclassicality tests as well as simple theoretical models.

	This paper is organized as follows.
	In Sec. \ref{sec:theory}, the theoretical model for our detection layout is elaborated and nonclassicality criteria are derived.
	The performed experiment is described in Sec. \ref{sec:experiment} with special emphasis on the used TESs.
	An extended analysis of our data, presented in Sec. \ref{sec:results}, includes the comparison of different forms of nonclassicality.
	We summarize and conclude in Sec. \ref{sec:summary}.

\section{Theory}\label{sec:theory}

\begin{figure}[b]
	\includegraphics[width=\columnwidth]{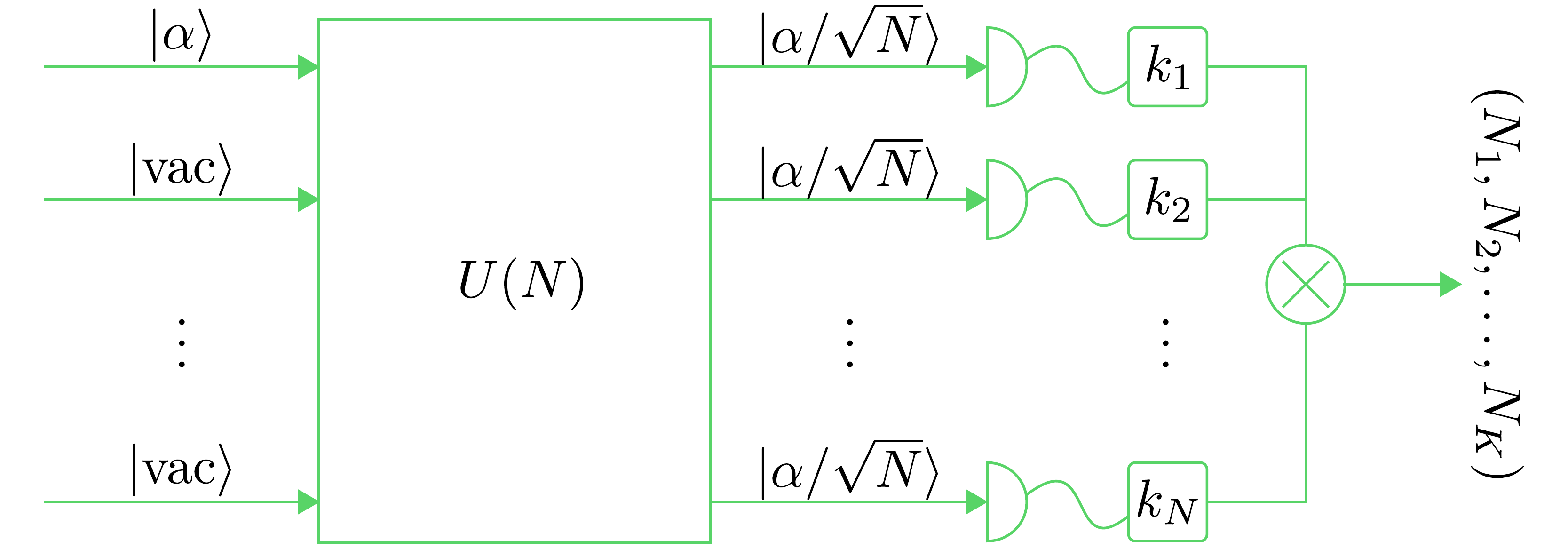}
	\caption{(Color online)
		Outline of the multiplexing scheme for a coherent state $|\alpha\rangle$.
		A balanced optical network---represented by the unitary $U(N)$---splits the incident coherent state $|\alpha\rangle$ into $|\alpha/\sqrt N\rangle^{\otimes N}$.
		The $n$th detector ($n\in\{1,\ldots,N\}$) gives an outcome $k_n\in\{0,\ldots,K\}$.
		The number of detectors which deliver the same, given outcome $k$ defines $N_k$.
	}\label{fig:multiplexing}
\end{figure}

	In this section, we derive the general, theoretical toolbox for describing the multiplexing arrangement with arbitrary detectors and for formulating the corresponding nonclassicality criteria.
	The measurement layout under study is shown in Fig. \ref{fig:multiplexing}.
	Our detection model shows that for any type of employed detector, the measured statistics can be described in the form of a quantum version of a multinomial statistics [Eq. \eqref{eq:clickcounting}].
	This leads to the formulation of nonclassicality criteria in terms of negativities in the normally ordered matrix of moments [Eq. \eqref{eq:fullMoM}].
	Especially, covariance-based criteria are discussed and related to previously known forms of nonclassicality.

\subsection{Preliminaries}

	We apply well-established concepts in quantum optics in this section.
	Namely, any quantum state of light $\hat\rho$ can be written in terms of the Glauber-Sudarshan representation \cite{G63,S63},
	\begin{align}\label{eq:GSrepresentation}
		\hat\rho=\int d^2\alpha\,P(\alpha)|\alpha\rangle\langle\alpha|.
	\end{align}
	From this diagonal expansion in terms of coherent states $|\alpha\rangle$, one observes that one can formulate the detection theory in terms of coherent states.
	A subsequent integration over the $P$ function then describes the model for any state.
	Furthermore, the definition of nonclassicality is also based on this representation.
	Namely, the state $\hat\rho$ is a classical state if and only if $P$ can be interpreted in terms of classical probability theory \cite{TG86,M86}, i.e., $P(\alpha)\geq0$.
	Whenever this cannot be done, $\hat\rho$ refers to a nonclassical state.

	Moreover, the $P$ function of a state is related to the normal ordering (denoted by ${:}\cdots{:}$) of measurement operators.
	For a detailed introduction to bosonic operator ordering, we refer to Ref. \cite{AW70}.
	It can be shown in general that any classical state obeys \cite{VW06}
	\begin{align}\label{eq:generalclassicalityconstraint}
		\langle{:}\hat f^\dag\hat f{:}\rangle\stackrel{\mathrm{cl.}}{\geq}0,
	\end{align}
	for any operator $\hat f$.
	In addition, we may recall that expectation values of normally ordered operators and coherent states can be simply computed by replacing the bosonic annihilation $\hat a$ and creation operator $\hat a^\dag$ with the coherent amplitude $\alpha$ and its complex conjugate $\alpha^\ast$, respectively.
	A violation of constraint \eqref{eq:generalclassicalityconstraint} necessarily identifies nonclassicality, which will be also used to formulate our nonclassicality criteria.

\subsection{Multiplexing detectors}

	The optical detection scheme under study, shown in Fig. \ref{fig:multiplexing}, consists of a balanced multiplexing network which splits a signal into $N$ modes.
	Those outputs are measured with $N$ identical detectors which can produce $K+1$ outcomes, labeled as $k=0,\ldots,K$.
	Let us stress that we make a clear distinction between the well-characterized optical multiplexing, the individual and unspecified detectors, and the resulting full detection scheme.

	In the multiplexing part, a coherent-state input $|\alpha\rangle$ is distributed over $N$ output modes.
	Further on, we have vacuum $|\mathrm{vac}\rangle=|0\rangle$ at all other $N-1$ input ports.
	In general, the $N$ input modes---defined by the bosonic annihilation operators $\hat a_{n,\mathrm{in}}$ ($\hat a_{1,\mathrm{in}}=\hat a$)---are transformed via the unitary $U(N)=(U_{m,n})_{m,n=1}^N$ into the output modes
	\begin{align}
		\hat a_{m,\mathrm{out}}=U_{m,1}\hat a_{1,\mathrm{in}}+\cdots+U_{m,N}\hat a_{N,\mathrm{in}}.
	\end{align}
	Taking the balanced splitting into account, it holds that $|U_{m,n}|=1/\sqrt N$.
	Adjusting the phases of the outputs properly, we get the following input-output relation
	\begin{align}\label{eq:coherentinout}
		|\alpha\rangle\otimes|0\rangle^{\otimes(N-1)}
		\stackrel{U(N)}{\longmapsto}|\alpha/\sqrt N\rangle^{\otimes N}.
	\end{align}
	Note that a balanced, but lossy network similarly yields $|\tau\alpha\rangle\otimes\cdots\otimes|\tau\alpha\rangle$ for $\tau\leq 1/\sqrt{N}$.

	For describing the detector, we do not make any specifications.
	Nevertheless, we will be able to formulate nonclassicality tests.
	The probability $p_k$ for the $k$th measurement outcome ($0\leq k\leq K$) for any type of detector can be written in terms of the expectation value of the POVM operators ${:}\hat\pi'_k{:}$, $p_k=\langle{:}\hat\pi'_k{:}\rangle$.
	Note that any operator can be written in a normally ordered form \cite{AW70,VW06} and that the POVM includes all imperfections of the individual detector, such as the quantum efficiency or nonlinear responses.
	For the coherent states $|\alpha/\sqrt{N}\rangle$, we have
	\begin{align}\label{eq:expectsingledetcoh}
		p_k(\alpha)=\langle\alpha/\sqrt{N}|{:}\hat\pi'_k{:}|\alpha/\sqrt{N}\rangle=\langle\alpha|{:}\hat\pi_k{:}|\alpha\rangle,
	\end{align}
	whereby we also define ${:}\hat\pi_k{:}$ in terms of ${:}\hat\pi'_k{:}$ through the mapping $\hat a\mapsto\hat a/\sqrt{N}$.

	We find that for a measurement with our $N$ detectors and our coherent output state \eqref{eq:coherentinout}, the probability to measure the outcome $k_n$ with the $n$th detector---more rigorously a coincidence $(k_1,\ldots,k_N)$ from the $N$ individual detectors---is given by
	\begin{align}
		&p_{k_1}(\alpha)
		\cdots
		p_{k_N}(\alpha)=\langle\alpha|{:}\hat\pi_{k_1}\cdots\hat\pi_{k_N}{:}|\alpha\rangle,
	\end{align}
	where we used the relation $\langle\alpha|{:}\hat A{:}|\alpha\rangle\langle\alpha|{:}\hat B{:}|\alpha\rangle=\langle\alpha|{:}\hat A\hat B{:}|\alpha\rangle$ for any two (or more) operators $\hat A$ and $\hat B$ and Eq. \eqref{eq:expectsingledetcoh}.
	The Glauber-Sudarshan representation \eqref{eq:GSrepresentation} allows one to write for any quantum state $\hat\rho$
	\begin{align}\label{eq:multimodenormal}
	\begin{aligned}
		p_{(k_1,\ldots,k_N)}=&\int d^2\alpha\,P(\alpha)p_{k_1}(\alpha)\cdots p_{k_N}(\alpha)
		\\
		=&\langle{:}\hat\pi_{k_1}\cdots\hat\pi_{k_N}{:}\rangle.
	\end{aligned}
	\end{align}

	So far we studied the individual parts, i.e., the optical multiplexing and the $N$ individual detectors, separately.
	To describe the full detection scheme in Fig. \ref{fig:multiplexing}, we need some additional combinatorics, which is fully presented in Appendix \ref{sec:detectionmodel}.
	There, the main idea is that one can group the individual detectors into subgroups of $N_k$ detectors which deliver the same outcome $k$.
	Suppose the individual detectors yield the outcomes $(k_1,\ldots,k_N)$.
	Then, $N_k$ is the number of individual detectors for which $k_n=k$ holds.
	In other words, $(N_0,\ldots,N_K)$ describes the coincidence that $N_0$ detectors yield the outcome $0$, $N_1$ detectors yield the outcome $1$, etc.
	Note that the total number of detectors is given by $N=N_0+\cdots+N_K$.

	The POVM representation $\hat\Pi_{(N_0,\dots,N_K)}$ for the event $(N_0,\ldots,N_K)$ is given in Eq. \eqref{eq:POVM}.
	In combination with Eq. \eqref{eq:multimodenormal}, we get for the detection layout in Fig. \ref{fig:multiplexing} the click-counting statistics of a state $\hat\rho$ as
	\begin{align}\label{eq:clickcounting}
	\begin{aligned}
		&c_{(N_0,\dots,N_K)}=\mathrm{tr}[\hat\rho\hat\Pi_{(N_0,\dots,N_K)}]
		\\=&\left\langle{:}\frac{N!}{N_0!\cdots N_K!}
			\hat\pi_0^{N_0}
			\cdots
			\hat\pi_K^{N_K}
		{:}\right\rangle,
	\end{aligned}
	\end{align}
	which is a normal-ordered version of a multinomial distribution.
	The click-counting statistics \eqref{eq:clickcounting} yields the probability that $N_0$ times the outcome $k=0$ together with $N_1$ times the outcome $k=1$, etc., is recorded with the $N$ individual detectors.
	Using Eq. \eqref{eq:GSrepresentation}, we can rewrite the click-counting distribution,
	\begin{align}
	\begin{aligned}
		\label{eq:clickcountingGS}
		c_{(N_0,\dots,N_K)}
		=&\int d^2\alpha\,P(\alpha)\frac{N!}{N_0!\cdots N_K!}
		\\&\times
		p_{0}(\alpha)^{N_0}
		\cdots
		p_{K}(\alpha)^{N_K}.
	\end{aligned}
	\end{align}
	In this form, we can directly observe that any classical statistics, $P(\alpha)\geq0$, is a classical average over multinomial probability distributions.

\subsection{Higher-order nonclassicality criteria}

	Our click-counting model \eqref{eq:clickcounting} describes a multiplexing scheme and applies to arbitrary detectors.
	One observes that its probability distribution is based on normally ordered expectation values of the form $\langle{:}\hat\pi_0^{m_0}\cdots\hat\pi_K^{m_K}{:}\rangle$.
	Hence, we can formulate nonclassicality criteria from inequality \eqref{eq:generalclassicalityconstraint} while expanding
	\begin{align}
		\hat f=\sum_{m_0+\cdots+m_K\leq N/2} f_{m_0,\ldots,m_K}\hat\pi_{0}^{m_0}\cdots\hat\pi_{K}^{m_K}.
	\end{align}
	This operator is chosen such that it solely includes the operators that are actually measured.
	We can write
	\begin{align}\label{eq:generalclassicalityconstraint1}
	\begin{aligned}
		\langle{:}\hat f^\dag\hat f{:}\rangle
		=&\sum_{\begin{smallmatrix}
			m_0+\cdots+m_K\leq N/2\\
			m'_0+\cdots+m'_K\leq N/2
		\end{smallmatrix}}
		f^\ast_{m_0,\ldots,m_K}
		\\&\times
		\left\langle{:}
			\hat\pi_0^{m_0+m'_0}\cdots\hat\pi_K^{m_K+m'_K}
		{:}\right\rangle
		f_{m'_0,\ldots,m'_K}
		\\
		=&\vec f^{\,\dag} M\vec f,
	\end{aligned}
	\end{align}
	with a vector $\vec f=(f_{m_0,\ldots,m_K})_{(m_0,\ldots,m_K)}$, using a multi-index notation, and the matrix of normally ordered moments $M$, which is defined in terms of the elements $\langle{:}\hat\pi_0^{m_0+m'_0} \cdots \hat\pi_K^{m_K+m'_K}{:}\rangle$.
	Also note that the order of the moments is bounded by the number of individual detectors, $N\geq m_0+\cdots+m_K+m'_0+\cdots+m'_K$, as the measured statistics \eqref{eq:clickcounting} only allows for retrieving them.

	As the non-negativity of the expression \eqref{eq:generalclassicalityconstraint1} holds for classical states [condition \eqref{eq:generalclassicalityconstraint}] and for all coefficients $\vec f$, we can equivalently write the following:
	A state is nonclassical if
	\begin{align}\label{eq:fullMoM}
		0\nleq M.
	\end{align}
	Conversely, the matrix of higher-order, normal-ordered moments $M$ is positive semidefinite for classical light.
	Note, it can be also shown (Appendix A in Ref. \cite{LSV15}) that the matrix of normally ordered moments can be equivalently expressed in a form that is based on central moments, $\langle{:}(\Delta\hat\pi_0)^{m_0+m_0'}\cdots(\Delta\hat\pi_K)^{m_K+m_K'}{:}\rangle$.

	For example and while restricting to the second-order submatrix, we get nonclassicality conditions in terms of normal-ordered covariances,
	\begin{align}\label{eq:secondordermatrix}
	\begin{aligned}
		0\nleq& M^{(2)}=\left(\langle{:}\Delta\hat\pi_k\Delta\hat\pi_{k'}{:}\rangle\right)_{k,k'=0,\ldots,K}
		\\
		&=\begin{pmatrix}
			\langle{:}(\Delta\hat\pi_0)^{2}{:}\rangle &\hdots& \langle{:}(\Delta\hat\pi_0)(\Delta\hat\pi_K){:}\rangle
			\\ \vdots&\ddots&\vdots \\
			\langle{:}(\Delta\hat\pi_0)(\Delta\hat\pi_K){:}\rangle &\hdots& \langle{:}(\Delta\hat\pi_K)^{2}{:}\rangle
		\end{pmatrix}.
	\end{aligned}
	\end{align}
	The relation $\langle{:}\hat\pi_K{:}\rangle=1-[\langle{:}\hat\pi_0{:}\rangle+\cdots+\langle{:}\hat\pi_{K-1}{:}\rangle]$ of general POVMs implies that the last row of $M^{(2)}$ is linearly dependent on the other ones.
	This further implies that zero is an eigenvalue of $M^{(2)}$.
	Hence, we get for any classical state that the minimal eigenvalue of this covariance matrix is necessarily zero.

	In order to relate our nonclassicality criteria to the measurement of the click-counting statistics \eqref{eq:clickcounting}, let us consider the generating function, which is given by
	\begin{align}
	\begin{aligned}
		&g(z_0,\ldots,z_N)=\overline{z_0^{N_0}\cdots z_K^{N_K}}
		\\
		=&\sum_{N_0+\cdots+N_K=N} c_{(N_0,\ldots,N_K)} z_0^{N_0}\cdots z_K^{N_K}
		\\
		=&\left\langle{:}
			\left(z_0\hat\pi_0+\cdots+z_K\hat\pi_K\right)^N
		{:}\right\rangle.
	\end{aligned}
	\end{align}
	The derivatives of the generating function relate the measured moments with the normally ordered ones,
	\begin{align}
		\nonumber &\left.\partial_{z_0}^{m_0}\cdots\partial_{z_K}^{m_K}g(z_0,\ldots,z_K)\right|_{z_0=\cdots=z_K=1}
		\\\nonumber =&\sum_{N_0+\cdots+N_K=N} c_{(N_0,\ldots,N_K)}\frac{N_0!}{(N_0-m_0)!}\cdots\frac{N_K!}{(N_K-m_K)!}
		\\\nonumber =&\overline{
			\left(N_0\right)_{m_0}
			\cdots
			\left(N_K\right)_{m_K}
		}
		\\=&\left(N\right)_{m_0+\cdots+m_K}
		\left\langle{:}
			\hat\pi_0^{m_0}\cdots\hat\pi_K^{m_K}
		{:}\right\rangle
		\label{eq:factorialmoments}
	\end{align}
	for $m_0+\cdots+m_K\leq N$ and $(x)_m=x(x-1)\cdots(x-m+1)=x!/(x-m)!$ being the falling factorial.
	Having a closer look at the second and third line of Eq. \eqref{eq:factorialmoments}, we see that the factorial moments $\overline{\left(N_0\right)_{m_0}\cdots\left(N_K\right)_{m_K}}$ can be directly sampled from $c_{(N_0,\ldots,N_K)}$.
	From the last two lines of Eq. \eqref{eq:factorialmoments} follows the relation to the normally ordered moments, which are needed for our nonclassicality tests.

\subsection{Second-order criteria}\label{subsec:nonclassicality2ndOrder}

	As an example and due to its importance, let us focus on the first- and second-order moments in detail.
	In addition, our experimental realization implements a single multiplexing step, $N=2$, which yields a restriction to second-order moments [see comment below Eq. \eqref{eq:generalclassicalityconstraint1}].
	As a special case of Eq. \eqref{eq:factorialmoments}, we obtain
	\begin{align}\label{eq:firstsecondsample}
		\langle{:}\hat\pi_{k}{:}\rangle=\frac{\overline{N_k}}{N}
		\text{ and }
		\langle{:}\hat\pi_k\hat\pi_{k'}{:}\rangle=\frac{\overline{N_kN_{k'}}-\delta_{k,k'}\overline{N_k}}{N(N-1)}
	\end{align}
	for $k,k'\in\{0,\ldots,K\}$.
	Hence, our covariances are alternatively represented by
	\begin{align}\label{eq:covariances}
		\langle{:}
			\Delta\hat\pi_{k}\Delta\hat\pi_{k'}
		{:}\rangle
		=\frac{
			N\overline{\Delta N_k\Delta N_{k'}}-\overline{N_k}\left(N\delta_{k,k'}-\overline{N_{k'}}\right)
		}{N^2(N-1)}.
	\end{align}
	As the corresponding matrix \eqref{eq:secondordermatrix} of normal-ordered moments is nonnegative for classical states, we get
	\begin{align}\label{eq:lettercriterion}
		0\stackrel{\mathrm{cl.}}{\leq}&N^2(N-1)M^{(2)}
		\\\nonumber&
		=\left(N\overline{\Delta N_k\Delta N_{k'}}-\overline{N_k}\left[N\delta_{k,k'}-\overline{N_{k'}}\right]\right)_{k,k'=0,\ldots,K}.
	\end{align}
	The violation of this specific constraint for classical states has been experimentally demonstrated for the generated quantum light \cite{Setal2017}.

	Let us consider other special cases of the general criterion.
	In particular, let us study the projections that result in a nonclassicality condition
	\begin{align}\label{eq:projectednonclassicality}
		\vec f^\mathrm{\,T} M^{(2)}\vec f<0,
	\end{align}
	see also Eqs. \eqref{eq:generalclassicalityconstraint} and \eqref{eq:secondordermatrix}.
	Note that $M^{(2)}$ is a real-valued and symmetric $(K+1)\times(K+1)$ matrix.
	Thus, it is sufficient to consider real-valued vectors $\vec f=(f_0,\ldots,f_K)^\mathrm{T}$.
	Further on, let us define the operator
	\begin{align}\label{eq:mudefinition}
		{:}\hat\mu{:}=f_0{:}\hat\pi_0{:}+\cdots+f_K{:}\hat\pi_K{:}.
	\end{align}
	Then, we can also read condition \eqref{eq:projectednonclassicality} as
	\begin{align}\label{eq:mucriterion}
		\langle{:}(\Delta\hat\mu)^2{:}\rangle<0.
	\end{align}
	That is, the fluctuations of the observable ${:}\hat\mu{:}$ are below those of any classical light field.
	In the following, we consider specific choices for $\vec f$ to formulate different nonclassicality criteria.

\subsubsection{Sub-multinomial light}\label{subsubsec:MultiLight}

	The minimization of \eqref{eq:projectednonclassicality} over all normalized vectors yields the minimal eigenvalue $Q_\mathrm{multi}$ of $M^{(2)}$.
	That is
	\begin{align}\label{eq:submultinomial}
		Q_\mathrm{multi}=\min_{\vec f:\vec f^\mathrm{\,T}\vec f=1}\vec f^\mathrm{\,T} M^{(2)}\vec f=\vec f_{0}^\mathrm{\,T}M^{(2)}\vec f_{0},
	\end{align}
	where $\vec f_{0}$ is a normalized eigenvector to the minimal eigenvalue.
	If we have $M^{(2)}\ngeq 0$, then we necessarily get $Q_\mathrm{multi}<0$.
	For classical states, we get $Q_\mathrm{multi}=0$; see the discussion below Eq. \eqref{eq:secondordermatrix}.
	As this criterion exploits the maximal negativity from covariances of the multinomial statistics, we refer to a radiation field with $Q_\mathrm{multi}<0$ as sub-multinomial light.

\subsubsection{Sub-binomial light}\label{subsubsec:BinLight}

	We can also consider the vector $\vec f=(0,1,\ldots,1)^\mathrm{T}$, which yields ${:}\hat\mu{:}=\hat 1-{:}\hat\pi_0{:}$.
	Hence, we have effectively reduced our system to a detection with a binary outcome, represented through the POVMs ${:}\hat\pi_0{:}$ and ${:}\hat\mu{:}=\hat 1-{:}\hat\pi_0{:}$.
	Using a proper scaling, we can write
	\begin{align}\label{eq:subbinomial}
	\begin{aligned}
		&\frac{(N-1)\vec f^{\,\mathrm T}M^{(2)}\vec f}{\langle{:}\hat\pi_0{:}\rangle(1-\langle{:}\hat\pi_0{:}\rangle)}
		=\frac{N\overline{(\Delta B)^2}-N\overline B+\overline B^2}{(N-\overline B)\overline{B}}
		\\=&N\frac{\overline{(\Delta B)^2}}{\overline{B}(N-\overline B)}-1=Q_\mathrm{bin},
	\end{aligned}
	\end{align}
	defining $B=N_1+\cdots+N_K=N-N_0$ and using Eq. \eqref{eq:firstsecondsample}.
	The condition $Q_\mathrm{bin}<0$ defines the notion of sub-binomial light \cite{SVA12} and is found to be a special case of inequality \eqref{eq:projectednonclassicality}.

\subsubsection{Sub-Poisson light}\label{subsubsec:PoisLight}

	Finally, we study criterion \eqref{eq:projectednonclassicality} for $\vec f=(0,1,\ldots,K)^\mathrm{T}$.
	We have ${:}\hat\mu{:}=\sum_{k=0}^K k {:}\hat\pi_k{:}$ and we also define
	\begin{align}
		A=\sum_{k=0}^K k N_k.
	\end{align}
	Their mean values are related to each other,
	\begin{align}
		\langle{:}\hat\mu{:}\rangle=\sum_{k=0}^K k\frac{\overline{N_k}}{N}=\frac{\overline{A}}{N}.
	\end{align}
	We point out that $\overline{N_k}/N$ can be also interpreted as probabilities, being nonnegative $\overline{N_k}/N\geq0$ and normalized $1=\overline{N_0}/N+\cdots+\overline{N_K}/N$ since $N=N_0+\cdots+N_K$.
	Further, we can write the normally ordered variance \eqref{eq:mucriterion} in the form
	\begin{align}
		&\langle{:}(\Delta\mu)^2{:}\rangle=\vec f^{\,\mathrm T}M^{(2)}\vec f
		\\=&\nonumber
		\frac{\overline{(\Delta A)^2}-\overline A}{N(N-1)}
		\\&-\frac{
			\left(\sum_{k=0}^K k^2\frac{\overline{N_k}}{N}\right)
			-\left(\sum_{k=0}^K k\frac{\overline{N_k}}{N}\right)^2
			-\left(\sum_{k=0}^K k\frac{\overline{N_k}}{N}\right)
		}{N-1}.\nonumber
	\end{align}
	Again, we can use a proper, nonnegative scaling to find
	\begin{align}\label{eq:subpoisson}
		\frac{\langle{:}(\Delta\mu)^2{:}\rangle}{\langle{:}\mu{:}\rangle}=&\frac{Q_\mathrm{Pois}-Q_\mathrm{Pois}'}{N-1},
		\\\nonumber\text{with }
		Q_\mathrm{Pois}=&\frac{\overline{(\Delta A)^2}}{\overline A}-1
		\\\nonumber\text{and }
		Q_\mathrm{Pois}'=&\frac{\left(
			\sum_{k=0}^K k^2\frac{\overline{N_k}}{N}\right)
			-\left(\sum_{k=0}^K k\frac{\overline{N_k}}{N}\right)^2
		}{
			\left(\sum_{k=0}^K k\frac{\overline{N_k}}{N}\right)
		}-1.
	\end{align}
	The parameters $Q_\mathrm{Pois}$ and $Q_\mathrm{Pois}^\prime$, often denoted as the Mandel or $Q$ parameter, relate to the notion of sub-Poisson light \cite{M79}.
	However, we have a difference of two such Mandel parameters in Eq. \eqref{eq:subpoisson}.
	The second parameter $Q_\mathrm{Pois}'$ can be considered as a correction, because the statistics of $A$ is only in a rough approximation a Poisson distribution.
	This is further analyzed in Appendix \ref{app:PoisPois}.

\subsection{Discussion}

	We derived the click-counting statistics \eqref{eq:clickcounting} for unspecified POVMs of the individual detectors.
	This was achieved by using the properties of a well-defined multiplexing scheme.
	We solely assumed that the $N$ detectors (with $K+1$ possible outcomes) are described by the same POVM.
	A deviation from this assumption can be treated as a systematic error; see Supplemental Material to Ref. \cite{Setal2017}.
	The full detection scheme was shown to result in a quantum version of multinomial statistics.
	This also holds true for an infinite, countable ($K=|\mathbb N|$) or uncountable ($K=|\mathbb R|$) set of outcomes, for which any measurement run can only deliver a finite sub-sample.
	For coherent light $|\alpha_0\rangle$, we get a true multinomial probability distribution; see Eq. \eqref{eq:clickcountingGS} for $P(\alpha)=\delta(\alpha-\alpha_0)$.
	For a binary outcome, $K+1=2$, we retrieve a binomial distribution \cite{SVA12a,SVA13}, which applies, for example, to avalanche photodiodes in the Geiger mode \cite{HSPGHNVS16,SBVHBAS15,SBDBJDVW16} or superconducting nanowire detectors \cite{BKSSV17,BKSSV17atm}.

	Further on, we derived higher-order nonclassicality tests which can be directly sampled from the data obtained from the measurement layout in Fig. \ref{fig:multiplexing}.
	Then, we focused on the second-order nonclassicality probes and compared the cases of sub-multinomial [Eq. \eqref{eq:submultinomial}], sub-binomial [Eq. \eqref{eq:subbinomial}], and (corrected) sub-Poisson [Eq. \eqref{eq:subpoisson}] light.
	The latter notion is related to nonclassicality in terms of photon-number correlation functions (see also Ref. \cite{ZM90}) and is a special case of our general criteria.
	Additionally, our method can be generalized to multiple multiplexing-detection arrangements to include multimode correlations similar to the approach in Ref. \cite{SVA13}.

	Recently, another interesting application was reported to characterize spatial properties of a beam profile with multipixel cameras \cite{CTFLMA16}.
	There, the photon-number distribution itself is described in terms of a multinomial statistics, and the Mandel parameter can be used to infer nonclassical light.
	Here, we show that an balanced multiplexing and any measurement POVM yield a click-counting statistics---describing a different statistical quantity than the photon statistics of a beam profile---in the form of a quantum version of a multinomial distribution leading to higher-order nonclassicality criteria.
	We also demonstrated that in some special scenarios (Appendix \ref{app:PoisPois}), a relation between the click statistic and photon statistics can be retrieved which is, however, much more involved in the general case; see also Sec. \ref{subsec:DetectorResponse}.

\section{Experiment}\label{sec:experiment}

	Before applying the derived techniques to our data, we describe the experiment and study some features of our individual detectors in this section.
	Especially, the response of our detectors is shown to have a nonlinear behavior which underlines the need for our nonclassicality criteria which are applicable to any type of detector.
	Additional details can be found in Appendix \ref{app:BinningCoincidences}.

\subsection{Setup description and characterization}

\begin{figure}[b]
	\includegraphics[width=\columnwidth]{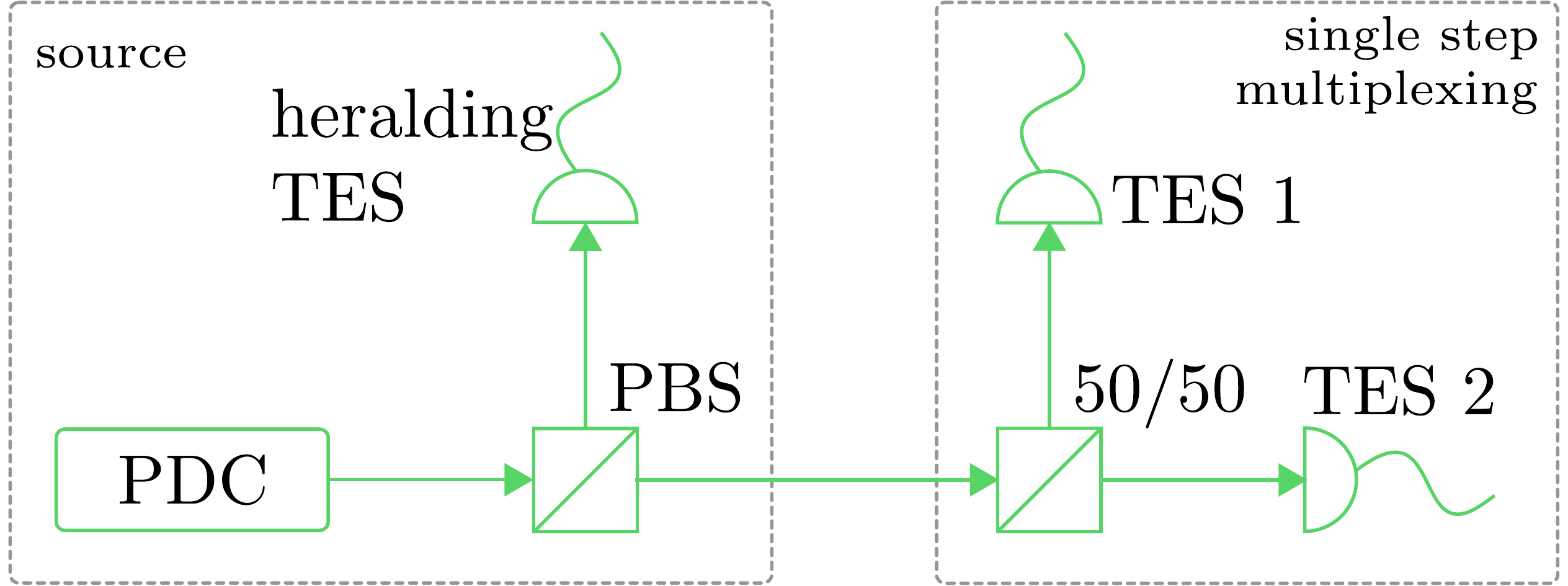}
	\caption{(Color online)
		Schematic setup.
		A parametric down-conversion (PDC) source emits photon pairs which are separated with a polarizing beam splitter (PBS).
		Conditioned on the measurement outcome of the heralding TES, different photon-number states are produced.
		A single multiplexing step is realized by splitting the photon states on a $50{/}50$ beam splitter and we detect them with two TESs.
	}\label{fig:setup}
\end{figure}

	An outline of our setup is given in Fig. \ref{fig:setup}.
	It is divided into a source that produces heralded photon states and a detection stage which represents one multiplexing step.
	In total, we use three superconducting TESs.
	For generating correlated photons, we employ a spontaneous parametric down-conversion (PDC) source.
	Here, we describe the individual parts in some more detail.

\subsubsection{The PDC source}

	Our spontaneous PDC source is a waveguide-written periodically poled potassium titanyl phosphate (PP-KTP) crystal which is $8\,\mathrm{mm}$ long.
	The type-II spontaneous PDC process is pumped with laser pulses at $775\,\mathrm{nm}$ and a full width at half maximum (FWHM) of $2\,\mathrm{nm}$ at a repetition rate of $75\,\mathrm{kHz}$.
	The heralding idler mode has a horizontal polarization and it is centered at $1554\,\mathrm{nm}$.
	The signal mode is vertically polarized and centered at $1546\,\mathrm{nm}$.
	A PBS spatially separates the output signal and idler pulses.
	An edge filter discards the pump beam.
	In addition, the signal and idler are filtered by $3\,\mathrm{nm}$ bandpass filters.
	This is done in order to filter out the broadband background which is typically generated in dielectric nonlinear waveguides \cite{ECMS11}.
	In general, such PDC sources have been proven to be well-understood and reliable sources of quantum light \cite{LCS09,KHQBSS13}.
	Hence, we may focus our attention on the employed detectors.

\subsubsection{The TES detectors}

	We use superconducting TESs as our photon detectors \cite{LMN08}.
	These TESs are micro-calorimeters consisting of $25\,\mathrm{\mu m}\times 25\,\mathrm{\mu m}\times 20\,\mathrm{nm}$ slabs of tungsten located inside an optical cavity with a gold backing mirror designed to maximize absorption at $1500\,\mathrm{nm}$.
	They are secured within a ceramic ferule as part of a self-aligning mounting system, so that the fiber core is well aligned to the center of the detector \cite{MLCVGN11}.
	The TESs are first cooled below their transition temperature within a dilution refrigerator and then heated back up to their transition temperature by Joule heating caused by a voltage bias, which is self-stabilized via an electro-thermal feedback effect \cite{I95}.
	Within this transition region, the steep resistance curve ensures that the small amount of heat deposited by photon absorption causes a measurable decrease in current flowing through the device.
	After photon absorption, the heat is then dissipated to the environment via a weak thermal link to the TES substrate.

	To read out the signal from this photon absorption process, the current change---produced by photon absorption in the TES---is inductively coupled to a superconducting quantum interference device (SQUID) module where it is amplified, and this signal is subsequently amplified at room temperature.
	This results in complex time-varying signals of about $5\,\mathrm{\mu s}$ duration.
	These signals are sent to a digitizer to perform fast analog-to-digital conversion, where the overlap with a reference signal is computed and then binned.
	This method allows us to process incoming signals at a speed of up to $100\,\mathrm{kHz}$.

	Our TESs are installed in a dilution refrigerator operating at a base temperature of about $70\,\mathrm{mK}$ and a cooling power of $400\,\mathrm{\mu W}$ at $100\,\mathrm{mK}$.
	One of the detectors has a measured detection efficiency of $0.98^{+ 0.02}_{- 0.08}$ \cite{HMGHLNNDKW15}.
	The other two TESs have identical efficiencies within the error of our estimation.

\subsection{Detector response analysis}\label{subsec:DetectorResponse}

	Even though we will not use specific detector characteristics for our analysis of nonclassicality, it is nevertheless scientifically interesting to study their response.
	This will also outline the complex behavior of superconducting detectors.
	For the time being, we ignore the detection events of the TESs 1 and 2 in Fig. \ref{fig:setup} and solely focus on the measurement of the heralding TES.

\begin{figure}[b]
	\includegraphics[width=\columnwidth]{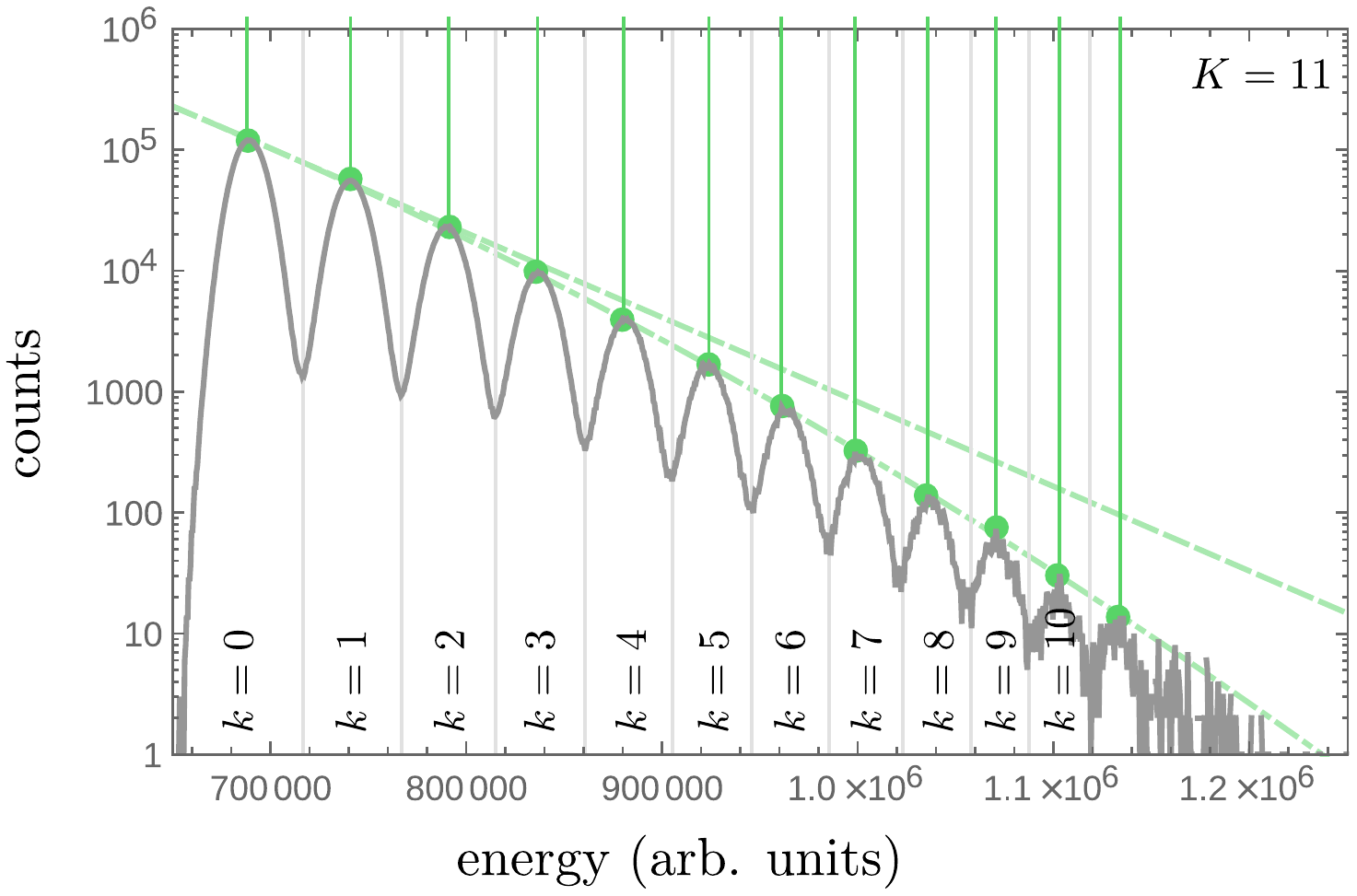}
	\caption{(Color online)
		The counts of the heralding TES (solid, gray curve); see also \cite{Setal2017}.
		Maxima for all $K+1=12$ intervals are shown as bullets.
		The dark vertical lines give the energy levels of the maxima.
		A nonlinear regression ($\log_{10} y=ax^2+bx+c$, dot-dashed line) and its tangent at the first maximum (dashed line) are additionally shown.
	}\label{fig:response}
\end{figure}

	In Fig. \ref{fig:response}, the measurement outcome of those marginal counts is shown.
	A separation into disjoint energy intervals represents our outcomes $k\in\{0,\ldots,11\}$ (see also Appendix \ref{app:BinningCoincidences}).
	The distribution around the peaked structures can be considered as fluctuations of the discrete energy levels (indicated by vertical dark green, solid lines).
	We observe that the difference between two discrete energies $E_n$ is not constant as one would expect from $E_{n+1}-E_n=\hbar\omega$, which will be discussed in the next paragraph.
	In addition, the marginal photon statistics should be given by a geometric distribution for the two-mode squeezed-vacuum state produced by our PDC source; see Appendix \ref{sec:theoreticalmodel}.
	In the logarithmic scaling in Fig. \ref{fig:response}, this would result in a linear function.
	However, we observe a deviation from such a model; compare light green, dashed and dot-dashed lines in Fig. \ref{fig:response}.

\begin{figure}[t]
	\includegraphics[width=\columnwidth]{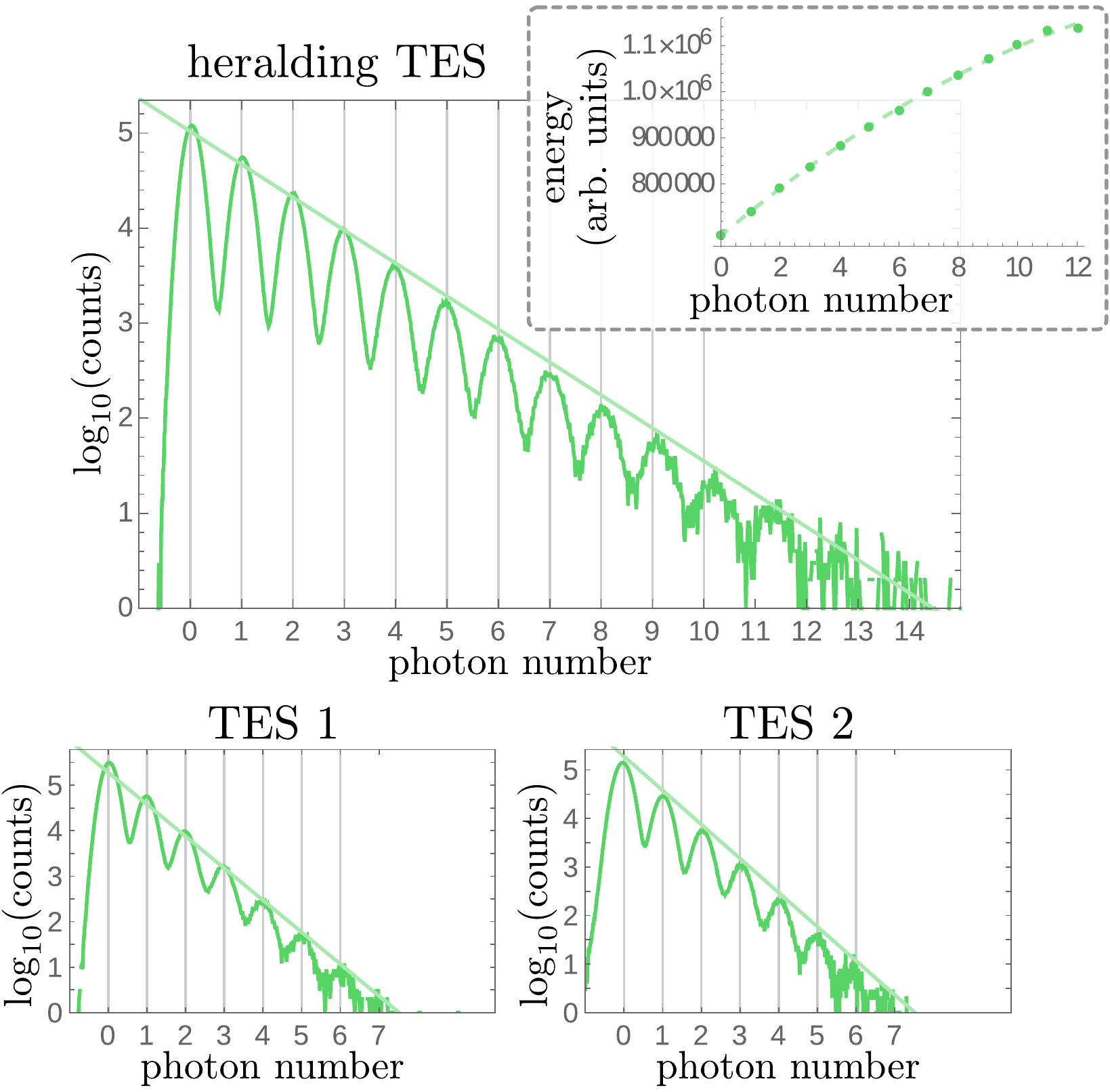}
	\caption{(Color online)
		A possible assignment between the measured counts and the photon-number estimate is shown for all TESs.
		As an example, the curve in the dashed box serves as the conversion from the measured energies of the heralding TES (points depict the maxima from Fig. \ref{fig:response}).
		This conversion yields an almost exponential ($\log_{10} y=ax+b$, light green lines) decay of the counts as it is expected for the geometric photon statistics produced by our source.
	}\label{fig:response-multi}
\end{figure}

	This deviation from the expected, linear behavior could have two origins:
	The source is not producing a two-mode squeezed-vacuum state (affecting the height of the peaks), or the detector, including the SQUID response, is not operating in a linear detection regime (influence on the horizontal axis).
	To counter the latter, the measured peak energies $E_n$---relating to the photon numbers $n$---have been fitted by a quadratic response function $n=aE_n^2+bE_n+c$; see the inset in Fig. \ref{fig:response-multi}.
	As a result of such a calibration, the peaked structure is well described by a linear function in $n$ for the heralding TES as shown in Fig. \ref{fig:response-multi} (top), which is now consistent with the theoretical expectation.
	The same nonlinear energy transformation also yields a linear $n$ dependence for the TESs 1 and 2 (cf. Fig. \ref{fig:response-multi}, bottom).
	Note that those two detectors only allow for a resolution of $K+1=8$ outcomes and that these two detectors have indeed a very similar response---the depicted linear function is identical for both.
	In conclusion, it is more likely that the measured nonlinear behavior in Fig. \ref{fig:response-multi} can be assigned to the detectors, and the PDC source is operating according to our expectations.

	In summary, we encountered an unexpected, nonlinear behavior of our data.
	To study this, a nonlinear fit was applied.
	This allowed us to make some predictions about the detector response in the particular interval of measurement while using known properties of our source.
	However, a lack of such extra knowledge prevents one from characterizing the detector.
	In Sec. \ref{sec:theory}, we have formulated nonclassicality tests which are robust against the particular response function of the individual detectors.
	They are accessible without any prior detector analysis and include the eventuality of nonlinear detector responses and other imperfections, such as quantum efficiency.
	With this general treatment, we also avoid the time-consuming detector tomography.

\section{Application}\label{sec:results}

	In this section, we apply the general theory, presented in Sec. \ref{sec:theory}, to our specific experimental arrangement, shown in Fig. \ref{fig:setup}.
	In the first step, we perform an analysis to identify nonclassicality which can be related to photon-number-based approaches.
	In the second step, we also compare the different criteria for sub-multinomial, sub-binomial, and sub-Poisson light for different realizations of our multi-photon states.

\subsection{Heralded multi-photon states}

	As derived in Appendix \ref{app:PoisPois}, the connection of the operator \eqref{eq:mudefinition}, for $f_k=k$, to the photon-number statistics for the idealized scenario of photoelectric detection POVMs is given by ${:}\hat\mu{:}=(\eta/N)\hat n$, where $\eta$ is the quantum efficiency of the individual detectors.
	This also relates---in this ideal case---the quantities
	\begin{align}\label{eq:mu_mheory}
		\langle{:}\hat\mu{:}\rangle=\frac{\eta}{N}\langle{:}\hat n{:}\rangle
		\text{ and }
		\langle{:}(\Delta\hat\mu)^2{:}\rangle=\frac{\eta^2}{N^2}\langle{:}(\Delta\hat n)^2{:}\rangle.
	\end{align}
	Recalling ${:}\hat n{:}=\hat n$, we see that $\langle{:}\hat\mu{:}\rangle$ is proportional to the mean photon number in this approximation.
	Similarly, we can connect $\langle{:}(\Delta\hat\mu)^2{:}\rangle$ to the normally ordered photon-number fluctuations.
	They are non-negative for classical states and negative for sub-Poisson light [see Eq. \eqref{eq:mucriterion}].

	An ideal PDC source is known to produce two-mode squeezed-vacuum states,
	\begin{align}\label{eq:tmsv}
		|q\rangle=\sqrt{1-|q|^2}\sum_{n=0}^\infty q^n |n\rangle\otimes|n\rangle,
	\end{align}
	where $|q|<1$.
	One mode can be used to produce multi-photon states by conditioning to the $l$th outcome of the heralding detector.
	Using photoelectric detector POVMs, we get the following mean value and the variances (Appendix \ref{sec:theoreticalmodel}):
	\begin{align}\label{eq:murelation}
	\begin{aligned}
		\langle{:}\hat\mu{:}\rangle{=}\frac{\eta}{N}\frac{\tilde\lambda+l}{1-\tilde\lambda}
		\text{ and }
		\langle{:}(\Delta \hat\mu)^2{:}\rangle{=}\frac{\eta^2}{N^2}\frac{(\tilde\lambda+l)^2-l(l+1)}{(1-\tilde\lambda)^2},
	\end{aligned}
	\end{align}
	with a transformed squeezing parameter $\tilde\lambda=(1-\tilde \eta)|q|^2$ and $\tilde \eta$ being the efficiency of the heralding detector.
	Note, we get the ideal $l$th Fock state, $|l\rangle$, for $\tilde\lambda\to0$.

\begin{figure}[b]
	\includegraphics[width=0.7\columnwidth]{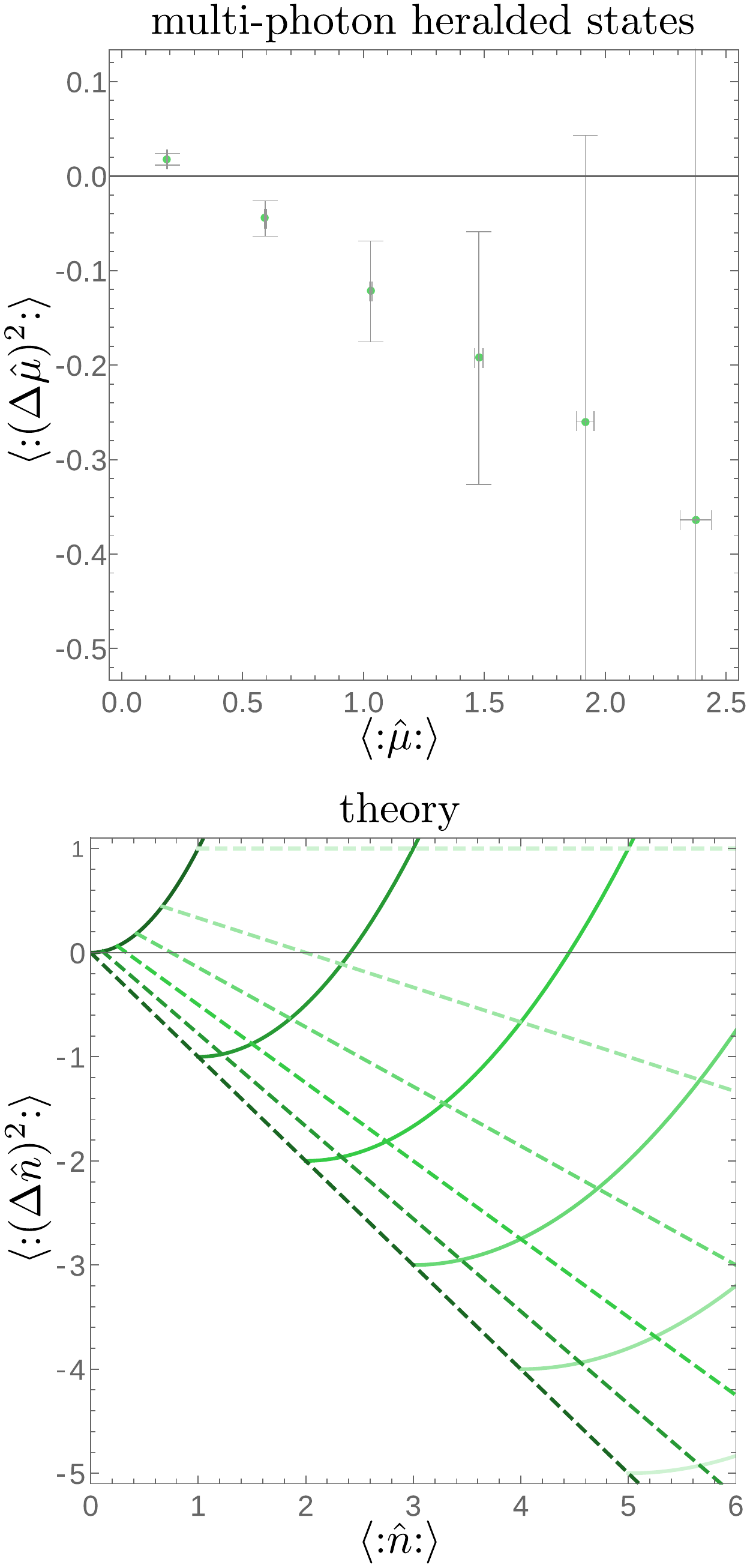}
	\caption{(Color online)
		The top panel shows the experimentally determined and normal-ordered mean value and variance of the operator $\hat\mu$ for the $l$th heralded state, increasing $l$ from left to right and $l=0,\ldots,5$.
		In an ideal case, those quantities relate to the photon-number statistics, cf. Eq. \eqref{eq:mu_mheory}.
		The bottom plot shows the theoretical expectations \eqref{eq:murelation} for photon number.
		The solid quadratic curves show the dependence for varying $\tilde\lambda$ and fixed $0\leq l\leq 5$ (lighter for increasing $l$).
		The dashed linear curves show the dependence for varying $l$ and fixed $\tilde\lambda \in\{0, 0.1,\ldots, 0.5\}$ (lighter for increasing $\tilde\lambda$).
	}\label{fig:mean-vs-var}
\end{figure}

\begin{figure*}
	\includegraphics[width=\textwidth]{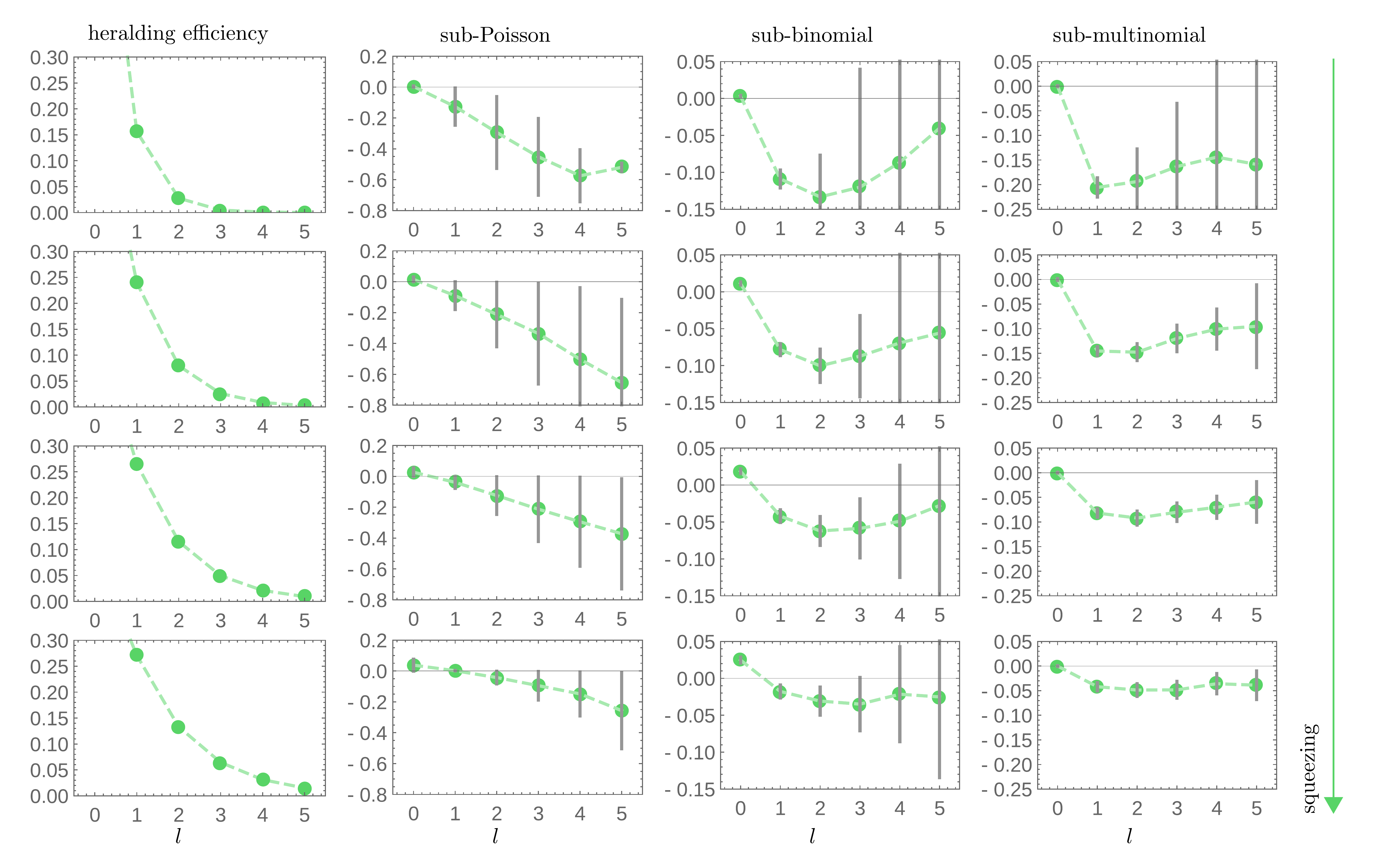}
	\caption{(Color online)
		Results of our analysis for different pump powers (indicated by an increasing squeezing from top to bottom rows) as functions of the heralding to $l$.
		The first column shows the success rate [Eq. \eqref{eq:gen}] for generating the $l$th multi-photon state.
		The second, third, and fourth columns depict the sub-Poisson, sub-binomial, and sub-multinomial nonclassicality criteria in Eqs. \eqref{eq:Poiss}, \eqref{eq:bin}, and \eqref{eq:multi}, respectively.
		For a better overview, dashed lines connect the individual data points.
	}\label{fig:results-multi}
\end{figure*}

	The experimental result is shown in the top panel of Fig. \ref{fig:mean-vs-var}.
	Using Eq. \eqref{eq:firstsecondsample}, we directly sampled the mean value and the variance of ${:}\hat\mu{:}=0{:}\hat\pi_0{:}+\cdots+K{:}\hat\pi_K{:}$ from the measured statistics for a heralding with $l=0,\ldots,5$.
	In this plot, $l$ increases from left to right relating to the increased mean photon numbers (including attenuations) of the heralded multi-photon states.
	The idealized theoretical modeling [see Eq. \eqref{eq:murelation}] is shown in the bottom part of Fig. \ref{fig:mean-vs-var}.
	Note, details of the error analysis have been formulated previously in Ref. \cite{Setal2017}.

	From the variances, we observe no nonclassicality when heralding to the $0$th outcome, which is expected as we condition on vacuum.
	In contrast, we can infer nonclassicality for the conditioning to higher outcomes of the heralding TES, $\langle{:}(\Delta\mu)^2{:}\rangle<0$ for $l>0$.
	We have a linear relation between the normally ordered mean and variance of ${:}\hat\mu{:}$, which is consistent with the theoretical prediction in Eq. \eqref{eq:murelation}.
	In the ideal case, the normal-ordered variance of the photon number for Fock states also decreases linearly with increasing $l$, $\langle l|\hat n|l\rangle=l$ and $\langle l|{:}(\Delta\hat n)^2{:}|l\rangle\rangle=-l$.
	It is also obvious that the errors are quite large for the verification of nonclassicality with this particular test for sub-Poisson light.
	We will discuss this in more detail in the next subsection.

\subsection{Varying pump power}\label{sec:VarPump}

	So far, we have studied measurements for a single pump power of the PDC process.
	However, the purity of the heralded states depends on the squeezing parameter, which is a function of the pump power.
	For instance, in the limit of a vanishing squeezing, we have the optimal approximation of the heralded state to a Fock state.
	However, the rate of the probabilistic generation converges to zero in the same limit (Appendix \ref{sec:theoreticalmodel}).
	Hence, we have additionally generated multi-photon states for different squeezing levels.
	The results of our analysis are shown in Fig. \ref{fig:results-multi} and will be discussed in the following.

	Suppose we measure the counts $C_l$ for the $l$th outcome of the heralding TES.
	The efficiency of generating this $l$th heralded state reads
	\begin{align}\label{eq:gen}
		\eta_\mathrm{gen}=\frac{C_l}{\sum_l C_l}.
	\end{align}
	From the model in Appendix \ref{sec:theoreticalmodel}, we expect that
	\begin{align}
		\eta_\mathrm{gen}=\frac{1-|q|^2}{1-|q|^2(1-\tilde\eta)}\left(\frac{\tilde\eta|q|^2}{1-|q|^2(1-\tilde\eta)}\right)^l.
	\end{align}
	The efficiency decays exponentially with $l$ and the decay is stronger for smaller squeezing or pump power---i.e., a decreasing $|q|^2$.
	In the left column of Fig. \ref{fig:results-multi}, we can observe this behavior.
	It can be seen in all other parts of Fig. \ref{fig:results-multi} that $\eta_\mathrm{gen}$ influences the significance of our results.
	A smaller $\eta_\mathrm{gen}$ value naturally implies a larger error because of a decreased sample size $C_l$.
	This holds for increasing $l$ and for decreasing squeezing.

	In the second column in Fig. \ref{fig:results-multi}, labeled as ``sub-Poisson'', we study the nonclassicality criterion
	\begin{align}\label{eq:Poiss}
		0>N^2(N-1)\vec f^{\,\mathrm T}M^{(2)}\vec f,
		\text{ for }
		\vec f=(0,1,\ldots, K)^\mathrm{T},
	\end{align}
	$N=2$, and $K=7$, which is related to sub-Poisson light (Sec. \ref{subsubsec:PoisLight}).
	The third column in Fig. \ref{fig:results-multi} correspondingly shows ``sub-binomial'' light (Sec. \ref{subsubsec:BinLight}),
	\begin{align}\label{eq:bin}
		0>N^2(N-1)\vec f^{\,\mathrm T}M^{(2)}\vec f
		\text{ for }
		\vec f=(0,1,\ldots, 1)^\mathrm{T}.
	\end{align}
	The last column, ``sub-multinomial'', depicts the nonclassicality criterion
	\begin{align}\label{eq:multi}
		0>&N^2(N-1)\vec f_0^{\,\mathrm T}M^{(2)}\vec f_0,
	\end{align}
	where $\vec f_0$ is a normalized eigenvector to the minimal eigenvalue of $M^{(2)}$ (Sec. \ref{subsubsec:MultiLight}).

	For all notations of nonclassicality under study, the heralding to the $0$th outcome is consistent with our expectation of a classical state, which also confirms that no fake nonclassicality is detected.
	For instance, applying the Mandel parameter to the data of this $0$th heralded stated without the corrections derived here [Eq. \eqref{eq:subpoisson}], we would observe a negative value; see also similar discussions in Refs. \cite{SVA12a,SBVHBAS15}.
	The case of a Poisson or binomial statistics tends to be above zero, whereas the multinomial case is consistent with the value of zero.
	This expectation has been justified below Eq. \eqref{eq:secondordermatrix}.

	A lot of information on the quantum-optical properties of the generated multi-photon ($l>0$) light fields can be concluded from Fig. \ref{fig:results-multi}.
	Let us mention some of them by focusing on a comparison.
	We have the trend that the notion of sub-Poisson light has the least significant nonclassicality.
	This is due to the vector $\vec f$ [Eq. \eqref{eq:Poiss}], which assigns a higher contribution to the larger outcome numbers.
	However, those contributions have lower count numbers, which consequently decreases the statistical significance.
	As depicted in Fig. \ref{fig:results-multi}, this effect is not present for sub-binomial light, which is described by a more or less balanced weighting of the different counts; see vector $\vec f$ in Eq. \eqref{eq:bin}.
	Still, this vector is fixed.

	The optimal vector is naturally computed by the sub-multinomial criterion in Eq. \eqref{eq:multi}.
	The quality of the verified nonclassicality is much better than for the other two scenarios of sub-Poisson and sub-binomial light in most of the cases.
	Let us mention that the normalized eigenvector to the minimal eigenvalue of the sampled matrix $M^{(2)}$ typically, but not necessarily, yields the minimal propagated error.
	Additionally, a lower squeezing level allows for the heralding of a state which is closer to an ideal Fock state.
	This results in higher negativities for decreasing squeezing and fixed outcomes $l$ in Fig. \ref{fig:results-multi}.
	However, the heralding efficiency $\eta_\mathrm{gen}$ is also reduced, which results in a larger error.

	Finally, we may point out that this comparative analysis of sub-Poisson, sub-binomial, and sub-multinomial light from data of a single detection arrangement would not be possible without the technique that has been elaborated in this paper (Sec. \ref{sec:theory}).

\section{Summary}\label{sec:summary}

	In summary, we constructed the quantum-optical framework to describe multiplexing schemes that employs arbitrary detectors and to verify nonclassicality of generated multi-photon states.
	We formulated the theory of such a detection layout together with nonclassicality tests.
	Further, we set up an experimental realization and applied our technique to the data.

	In a first step, the theory was formulated.
	We proved that the measured click-counting statistics of the scheme under study is always described by a quantum version of the multinomial statistics.
	In fact, for classical light, this probability distribution can be considered as a mixture of multinomial statistics.
	This bounds the minimal amount of fluctuations which can be observed for classical radiation fields.
	More precisely, the matrix of higher-order, normally ordered moments, which can be directly sampled from data, can exhibit negative eigenvalues for nonclassical light.
	As a particular example, we discussed nonclassicality tests based on the second-order covariance matrix, which led to establishing the concept of sub-multinomial light.
	Previously studied notions of nonclassicality, i.e, sub-Poisson and sub-binomial light, have been found to be special cases of our general nonclassicality criteria.

	In our second part, the experiment was analyzed.
	Our source produces correlated photon pairs by a parametric-down-conversion process.
	A heralding to the outcome of a detection of the idler photons with a transition-edge sensor produced multi-photon states in the signal beam.
	A single multiplexing step was implemented with a subsequent detection by two transition-edge sensors to probe the signal field.
	The complex function of these detectors was discussed by demonstrating their nonlinear response to the number of incident photons.
	Consequently, without worrying about this unfavorable feature, we applied our robust nonclassicality criteria to our data.
	We verified the nonclassical character of the produced quantum light.
	The criterion of sub-multinomial light was shown to outperform its Poisson and binomial counterparts to the greatest possible extent.

	In conclusion, we presented a detailed and more extended study of our approach in Ref. \cite{Setal2017}.
	We formulated the general positive-operator-valued measure and generalized the nonclassicality tests to include higher-order correlations which become more and more accessible with an increasing number of multiplexing steps.
	In addition, details of our data analysis and a simple theoretical model were considered.
	Thus, we described a robust detection scheme to verify quantum correlations with unspecified detectors and without introducing fake nonclassicality.

\begin{acknowledgments}
	The project leading to this application has received funding from the European Union's Horizon 2020 research and innovation programme under grant agreement No 665148.
	A.~E. is supported by EPSRC EP/K034480/1.
	J.~J.~R. is supported by the Netherlands Organization for Scientific Research (NWO).
	W.~S.~K is supported by EPSRC EP/M013243/1.
	S.~W.~N., A.~L., and T.~G. are supported by the Quantum Information Science Initiative (QISI).
	I.~A.~W. acknowledges an ERC Advanced Grant (MOQUACINO).
	The authors thank Johan Fopma for technical support.
	The authors gratefully acknowledge helpful comments by Tim Bartley and Omar Maga\~{n}a-Loaiza.

	Contributions of this work by NIST, an agency of the U.S. Government, are not subject to U.S. copyright.
\end{acknowledgments}

\appendix

\section{Combinatorics and POVM elements}\label{sec:detectionmodel}

	Here, we provide the algebra that is needed to get from Eq. \eqref{eq:multimodenormal} to Eq. \eqref{eq:clickcounting}.
	More rigorously, we use combinatorial methods to formulate the POVM $\hat\Pi_{(N_0,\ldots,N_K)}$ in terms of the POVM ${:}\hat\pi_{k_1}\cdots\hat\pi_{k_N}{:}$.
	Say $N_k$ is the number of elements of $(k_1,\ldots,k_N)$ which take the value $k$.
	Then, $(N_0,\ldots,N_K)$ describes the coincidence that $N_0$ detectors yield the outcome $0$, $N_1$ detectors yield the outcome $1$, etc.
	One specific and ordered measurement outcome is defined by $(k_{0,1},\ldots,k_{0,N})$, with
	\begin{align}\label{eq:specific_coincidence}
		k_{0,n}=\left\lbrace\begin{array}{lcc}
			0 &\text{ for }& 1\leq n \leq N_0,\\
			1 &\text{ for }& N_0+1\leq n \leq N_0+N_1,\\
			&\vdots&\\
			K &\text{ for }& N_0+\cdots+N_{K-1}+1\leq n \leq N,\\
		\end{array}\right.
	\end{align}
	which results in a given $(N_0,\ldots,N_K)$, where the total number of detectors is $N=N_0+\cdots+N_K$.
	This specific example can be used to represent all similar outcomes as we will show now.
	The $(k_1,\ldots,k_N)$ for the same combination $(N_0,\ldots,N_K)$ can be obtained from $(k_{0,\sigma(1)},\ldots,k_{0,\sigma(N)})$ via a permutation $\sigma\in\mathcal S_N$ of the elements.
	Here $\mathcal S_N$ denotes the permutation group of $N$ elements which has a cardinality of $N!$.
	Note that all permutations $\sigma$ which exchange identical outcomes result in the same tuple.
	This means for the outcome defined in Eq. \eqref{eq:specific_coincidence} that $(k_{0,\sigma(1)},\ldots,k_{0,\sigma(N)})=(k_{0,1},\ldots,k_{0,N})$ for any permutation of the form $\sigma\in\mathcal S_{N_0}\times\cdots\times\mathcal S_{N_K}$.
	Therefore, the POVM element for a given $(N_0,\ldots,N_K)$ can be obtained by summing over all permutations $\sigma\in\mathcal S_N$ of the POVMs of individual outcomes ${:}\hat\pi_{k_{0,1}}\cdots\hat\pi_{k_{0,N}}{:}$ [Eq. \eqref{eq:multimodenormal}] while correcting for the $N_0!\cdots N_K!$ multi-counts.
	More rigorously, we can write
	\begin{align}\label{eq:POVM}
		\nonumber
		\hat\Pi_{(N_0,\dots,N_K)}
		=&\frac{1}{N_0!\cdots N_K!}\sum_{\sigma\in\mathcal S_N}{:}\hat\pi_{k_{0,\sigma(1)}}\cdots\hat\pi_{k_{0,\sigma(N)}}{:}
		\\=&\frac{N!}{N_0!\cdots N_K!}{:}
			\hat\pi_0^{N_0}
			\cdots
			\hat\pi_K^{N_K}
		{:},
	\end{align}
	where relations of the form ${:}\hat A\hat B\hat A{:}={:}\hat A^2\hat B{:}$ have been used.

\section{Corrected Mandel parameter}\label{app:PoisPois}

	For the nonclassicality test in Sec. \ref{subsubsec:PoisLight}, we could assume a detector which can discriminate $K=\infty$ measurement outcomes, which are related to measurement operators of a Poisson form, ${:}\hat\pi_k'{:}={:}\hat \Gamma^ke^{-\hat\Gamma}{:}/k!$ \cite{KK64}, where $\hat\Gamma=\eta\hat n$ is an example of a linear detector response function ($\eta$ quantum efficiency).
	Using the definition \eqref{eq:expectsingledetcoh}, we get ${:}\hat\pi_k{:}={:}(\hat \Gamma/N)^ke^{-\hat\Gamma/N}{:}/k!$, where the denominator $N$ accounts for the splitting into $N$ modes \cite{SVA13}.
	This idealized model yields $\langle{:}\hat\mu{:}\rangle=\langle{:}(\hat\Gamma/N){:}\rangle$ and
	\begin{align}
	\begin{aligned}
		\sum_{k=0}^\infty k^2 \frac{\overline{N_k}}{N}=\frac{\langle{:}\hat\Gamma^2{:}\rangle}{N^2}{+}\frac{\langle{:}\hat\Gamma{:}\rangle}{N}
		\text{ and }
		\overline{A^2}= \langle{:}\hat\Gamma^2{:}\rangle{+}\langle{:}\hat\Gamma{:}\rangle.
	\end{aligned}
	\end{align}
	Hence, we have $Q_\mathrm{Pois}=\langle{:} (\Delta\hat\Gamma)^2{:}\rangle/\langle{:}\hat\Gamma{:}\rangle=NQ_\mathrm{Pois}'$ and
	\begin{align}
		\frac{\langle{:}(\Delta\mu)^2{:}\rangle}{\langle{:}\mu{:}\rangle}=\frac{1}{N}Q_\mathrm{Pois}=\frac{\eta}{N}\frac{\langle{:}(\Delta\hat n)^2{:}\rangle}{\langle{:}\hat n{:}\rangle}.
	\end{align}
	Thus, we have shown that for photoelectric detection models, we retrieve the notation of sub-Poisson light, $Q_\mathrm{Pois}<0$, from the general form \eqref{eq:subpoisson}, which includes a correction term.

\section{Binning and measured coincidences}\label{app:BinningCoincidences}

	The data in Fig. \ref{fig:response-multi} (Sec. \ref{sec:experiment}) are grouped in disjoint intervals around the peaks, representing the photon numbers.
	They define the outcomes $k=0,\ldots,K$.
	Because we are free in the choice of the intervals, we studied different scenarios and found that the given one is optimal from the information-theoretic perspective.
	On the one hand, if the current intervals are divided into smaller ones, we distribute the data of one photon number among several outcomes.
	This produces redundant information about this photon number.
	On the other hand, we have a loss of information about the individual photon numbers if the interval stretches over multiple photon numbers.
	This explains our binning as shown in Fig. \ref{fig:response-multi}.

\begin{figure}[ht]
	\includegraphics[width=\columnwidth]{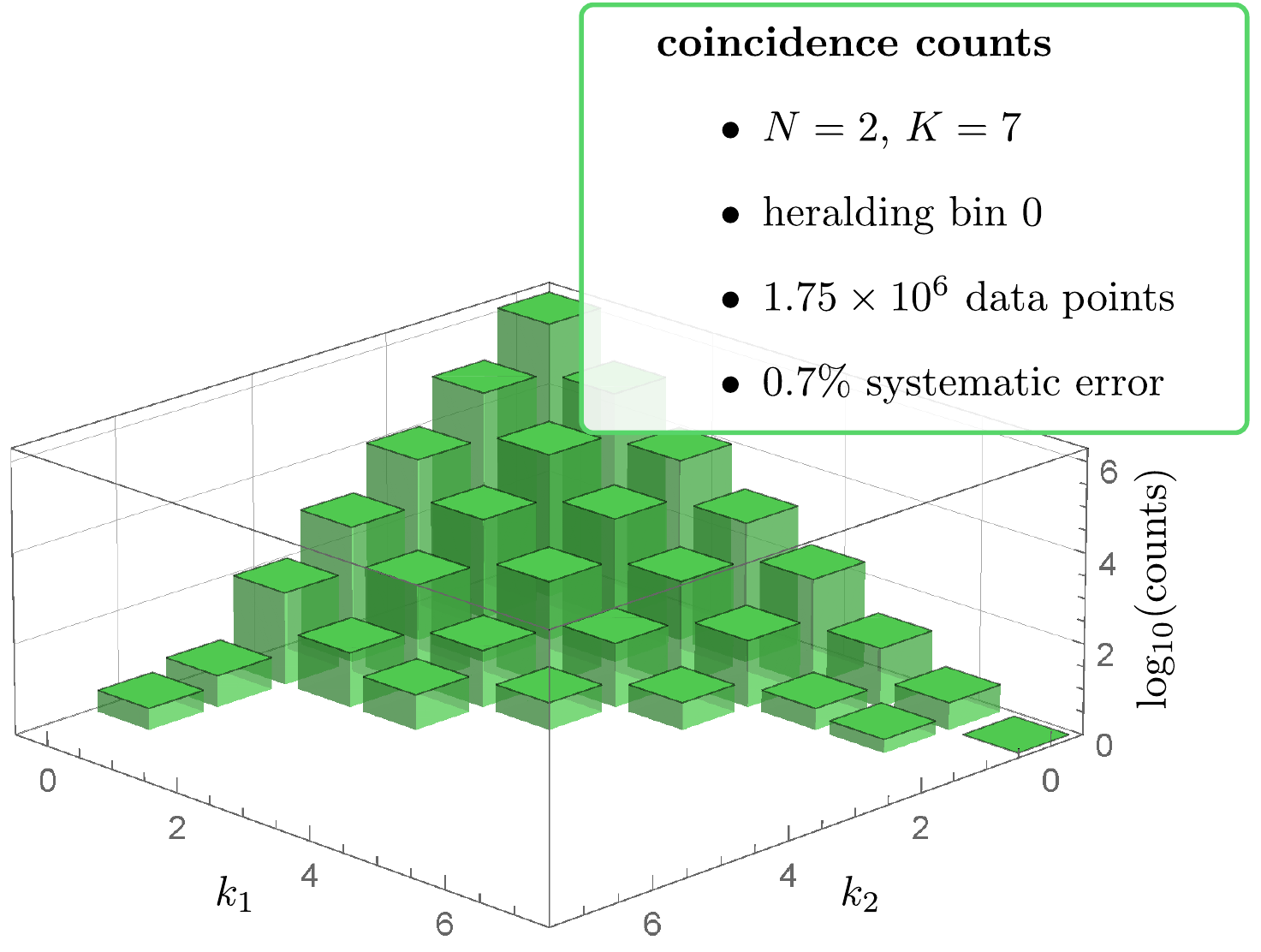}
	\caption{(Color online)
		Example of raw coincidence counts.
		$k_n$ is the outcome of the $n$th individual detector ($n=1,2$) for which the count was recorded.
		Some additional information on the statistics is given in the inset.
		The depicted state is produced by conditioning on the $0$th outcome of the heralding TES in Fig. \ref{fig:setup}.
	}\label{fig:coincidences}
\end{figure}

	An example of a measured coincidence statistics for outcomes $(k_1,k_2)$ is shown in Fig. \ref{fig:coincidences}.
	There, we consider a state which is produced by the simplest conditioning to the $0$th outcome of the heralding TES.

	Based on this plot, let us briefly explain how these coincidences for $(k_1,k_2)$ result in the statistics $c_{(N_0,\dots,N_K)}$ for $(N_0,\ldots,N_K)$ and $K=7$.
	The counts on the diagonal, $k_1=k_2=k$, of the plot yield $c_{(N_0,\dots,N_K)}$ for $N_k=2$ and $N_{k'}=0$ for $k'\neq k$.
	For example, the highest counts are recorded for $(k_1,k_2)=(0,0)$ in Fig. \ref{fig:coincidences} which gives $c_{(2,0,\ldots,0)}$ when normalized to all counts.
	Off-diagonal combinations, $k_1\neq k_2$, result in $c_{(N_0,\dots,N_K)}$ for $N_{k_1}=N_{k_2}=1$ and $N_k=0$ otherwise.
	For example, the normalized sum of the counts for $(k_1,k_2)\in\{(0,1),(1,0)\}$ yields $c_{(1,1,0,\ldots,0)}$.
	As we have $N=2$ TESs in our multiplexing scheme and $N_0+\cdots+N_K=N$, the cases $k_1=k_2$ and $k_1\neq k_2$ already define the full distribution $c_{(N_0,\dots,N_K)}$.

	The asymmetry in the counting statistics between the two detectors results in a small systematic error $\lesssim1\%$.
	One should keep in mind that the counts are plotted in a logarithmic scale.
	For all other measurements of heralded multi-photon states, this error is in the same order \cite{Setal2017}.

\section{Simplified theoretical model}\label{sec:theoreticalmodel}

	Let us analytically compute the quantities which are used for the simplified description of physical system under study.
	The PDC source produces a two-mode squeezed-vacuum state \eqref{eq:tmsv}, where the first mode is the signal and the second mode is the idler or herald.
	In our idealized model, the heralding detector is supposed to be a photon-number-resolving detector with a quantum efficiency $\tilde \eta$.
	A multiplexing and a subsequent measurement with $N$ photon-number-resolving detectors ($K=\infty$) are employed for the click counting.
	Each of the photon-number-resolving detector's POVM elements is described by
	\begin{align}
		{:}\hat\pi_k{:}={:}\frac{(\eta\hat n/N)^k}{k!}e^{-\eta\hat n/N}{:}.
	\end{align}
	In addition, we will make use of the relations ${:}e^{y\hat n}{:}=(1+y)^{\hat n}$ (cf., e.g., Ref. \cite{SVA14}) and
	\begin{align}
	\begin{aligned}
		\partial_z^k{:}e^{[z-1]y\hat n}{:}|_{z=1}=&{:}(y\hat n)^k{:},
		\\
		\frac{1}{k!}\partial_z^k{:}e^{[z-1]y\hat n}{:}|_{z=0}=&{:}\frac{(y\hat n)^k}{k!}e^{-y\hat n}{:}.
	\end{aligned}
	\end{align}

	For this model, we can conclude that the two-mode generating function for the considered two-mode squeezed-vacuum state reads
	\begin{align}
	\begin{aligned}
		&\Gamma(z,\vec x)=\langle{:}e^{[z-1]\tilde\eta\hat n}\otimes e^{[\|\vec x\|-1]\eta\hat n/N}{:}\rangle
		\\=&\frac{1-|q|^2}{1-|q|^2(1-\tilde\eta+\tilde\eta z)(1-\eta+\eta \|\vec x\|/N)}.
	\end{aligned}
	\end{align}
	where $z\in [0,1]$ relates the heralding mode and the components of $\vec x\in[0,1]^N$ (recall that $\|\vec x\|=\sum_n x_n$) to the outcomes of the $N$ detectors in the multiplexing scheme.
	From this generating function, we directly deduce the different properties that are used in this paper for comparing the measurement with our model.
	The needed derivatives are
	\begin{align}
	\begin{aligned}
		&\partial_{\vec x}^{\vec k}\partial_z^l\Gamma(z,\vec x)=\partial_{\|\vec x\|}^{k}\partial_z^l\Gamma(z,\vec x)
		\\=&(1-|q|^2)l!k!\frac{
			\left[(\eta/N)|q|^2z'\right]^k
			\left[\tilde \eta|q|^2x'\right]^l
		}{\left[1-|q|^2x'z'\right]^{k+l+1}}
		\\&\times\sum_{j=0}^{\min\{k,l\}}\frac{(k+l-j)!}{j!(k+j)!(l-j)!}\left[
			\frac{1-|q|^2x'z'}{|q|^2x'z'}
		\right]^j,
	\end{aligned}
	\end{align}
	where $k=\|\vec k\|$, $x'=1-\eta +\eta\|\vec x\|/N$, and $z'=1-\tilde\eta +\tilde\eta z$.
	It is also worth mentioning that the case $N=1$ yields the result for photon-number-resolving detection without multiplexing.

	The marginal statistics of the heralding detector reads
	\begin{align}\nonumber
		&\tilde p_l=\frac{1}{l!}\partial_z^l\Gamma(z,\vec x)|_{z=0,x_1=\cdots=x_N=1}
		\\
		=&\frac{1-|q|^2}{1-|q|^2(1-\tilde\eta)}\left(\frac{\tilde\eta|q|^2}{1-|q|^2(1-\tilde\eta)}\right)^l.
	\end{align}
	The marginal statistics of the $n$th detector is
	\begin{align}\nonumber
		&\frac{1}{k_n!}\partial_{x_n}^{k_n}\Gamma(1,\vec x)|_{x_n=0,z=1=x_1=\cdots=x_{n-1}=x_{n+1}=\cdots=x_N}
		\\=&\frac{1-|q|^2}{1-|q|^2(1-\eta/N)}\left(\frac{\eta|q|^2/N}{1-|q|^2(1-\eta/N)}\right)^{k_n}.
	\end{align}
	In addition, the case of no multiplexing ($N=1$ and $x\cong\vec x$) yields for the $l$th heralded state the following first and second normally ordered photon numbers:
	\begin{align}
		\langle{:}(\eta\hat n){:}\rangle=&\frac{1}{\tilde p_ll!}\partial_x\partial_z^l\Gamma(z,x)|_{z=0,x=1}
		=\eta\frac{l+\tilde\lambda}{1-\tilde\lambda},
		\\\nonumber
		\langle{:}(\eta\hat n)^2{:}\rangle=&\frac{1}{\tilde p_ll!}\partial_x^2\partial_z^l\Gamma(z,x)|_{z=0,x=1}
		\\=&\eta^2\frac{(2(l+\tilde\lambda)^2-l(l+1))}{(1-\tilde\lambda)^2},
	\end{align}
	with $\tilde\lambda=(1-\tilde\eta)|q|^2$.
	The corresponding photon distribution (i.e., for $\eta=1$) of the $l$th multi-photon state reads
	\begin{align}
	\begin{aligned}
		\tilde p_{k|l}=&\frac{1}{\tilde p_l}\frac{1}{k!l!}\partial_x^k\partial_z^l \Gamma(z,x)|_{z=x=0}
		\\=&\left\lbrace\begin{array}{lcc}
			0 &\text{ for }& k<l,\\
			\binom{k}{l}(1-\tilde\lambda)^{l+1}\tilde\lambda^{n-l} &\text{ for }& k\geq l.
		\end{array}\right.
	\end{aligned}
	\end{align}
	For $\tilde\lambda\to0$, we have $\tilde p_{k|l}=\delta_{k,l}$, which is the photon statistics of the $l$th Fock state.



\begin{thebibliography}{99}
	\bibitem{Setal2017}
		J. Sperling, W. R. Clements, A. Eckstein, M. Moore, J. J. Renema, W. S. Kolthammer, S. W. Nam, A. Lita, T. Gerrits, W. Vogel, G. S. Agarwal, and I. A. Walmsley,
		Detector-Independent Verification of Quantum Light,
		\href{https://doi.org/10.1103/PhysRevLett.118.163602}{Phys. Rev. Lett. \textbf{118}, 163602 (2017)}.
	\bibitem{E05}
		A. Einstein,
		\"{U}ber einen die Erzeugung und Verwandlung des Lichtes betreffenden heuristischen Gesichtspunkt,
		\href{http://dx.doi.org/10.1002/andp.19053220607}{Ann. Phys. (Leipzig) \textbf{17}, 132 (1905)}.
	\bibitem{BC10}
		G. S. Buller and R. J. Collins,
		Single-photon generation and detection,
		\href{http://dx.doi.org/10.1088/0957-0233/21/1/012002}{Meas. Sci. Technol. \textbf{21}, 012002 (2010)}.
	\bibitem{CDKMS14}
		C. J. Chunnilall, I. P. Degiovanni, S. K\"{u}ck, I. M\"{u}ller, and A. G. Sinclair,
		Metrology of single-photon sources and detectors: A review,
		\href{http://dx.doi.org/10.1117/1.OE.53.8.081910}{Opt. Eng. \textbf{53}, 081910 (2014)}.
	\bibitem{GT07}
		N. Gisin and R. Thew,
		Quantum communication,
		\href{https://doi.org/10.1038/nphoton.2007.22}{Nat. Photon. \textbf{1}, 165 (2007)}.
	\bibitem{S09}
		J. H. Shapiro,
		The Quantum Theory of Optical Communications,
		\href{http://dx.doi.org/10.1109/JSTQE.2009.2024959}{IEEE J. Sel. Top. Quantum Electron. \textbf{15}, 1547 (2009)}.
	\bibitem{SV11}
		A. A. Semenov and W. Vogel,
		Fake violations of the quantum Bell-parameter bound,
		\href{https://doi.org/10.1103/PhysRevA.83.032119}{Phys. Rev. A \textbf{83}, 032119 (2011)}.
	\bibitem{GLLSSMK11}
		I. Gerhardt, Q. Liu, A. Lamas-Linares, J. Skaar, V. Scarani, V. Makarov, and C. Kurtsiefer,
		Experimentally Faking the Violation of Bell’s Inequalities,
		\href{https://doi.org/10.1103/PhysRevLett.107.170404}{Phys. Rev. Lett. \textbf{107}, 170404 (2011)}.
	\bibitem{SVA12a}
		J. Sperling, W. Vogel, and G. S. Agarwal,
		True photocounting statistics of multiple on-off detectors,
		\href{http://dx.doi.org/10.1103/PhysRevA.85.023820}{Phys. Rev. A \textbf{85}, 023820 (2012)}.
	\bibitem{S63}
		E. C. G. Sudarshan,
		Equivalence of Semiclassical and Quantum Mechanical Descriptions of Statistical Light Beams,
		\href{http://dx.doi.org/10.1103/PhysRevLett.10.277}{Phys. Rev. Lett. \textbf{10}, 277 (1963)}.
	\bibitem{G63}
		R. J. Glauber,
		Coherent and incoherent states of the radiation field,
		\href{http://dx.doi.org/10.1103/PhysRev.131.2766}{Phys. Rev. \textbf{131}, 2766 (1963)}.
	\bibitem{TG86}
		U. M. Titulaer and R. J. Glauber,
		Correlation functions for coherent fields,
		\href{http://dx.doi.org/10.1103/PhysRev.140.B676}{Phys. Rev. \textbf{140}, B676 (1965)}.
	\bibitem{M86}
		L. Mandel,
		Non-classical states of the electromagnetic field,
		\href{http://dx.doi.org/10.1088/0031-8949/1986/T12/005}{Phys. Scr. \textbf{T12}, 34 (1986)}.
	\bibitem{MBWLN10}
		A. Miranowicz, M. Bartkowiak, X. Wang, Yu-xi Liu, and F. Nori,
		Testing nonclassicality in multimode fields: A unified derivation of classical inequalities,
		\href{http://dx.doi.org/10.1103/PhysRevA.82.013824}{Phys. Rev. A \textbf{82}, 013824 (2010)}.
	\bibitem{SRV05}
		E. Shchukin, Th. Richter, and W. Vogel,
		Nonclassicality criteria in terms of moments,
		\href{http://dx.doi.org/10.1103/PhysRevA.71.011802}{Phys. Rev. A \textbf{71}, 011802(R) (2005)}.
	\bibitem{M79}
		L. Mandel,
		Sub-Poissonian photon statistics in resonance fluorescence,
		\href{http://dx.doi.org/10.1364/OL.4.000205}{Opt. Lett. \textbf{4}, 205 (1979)}.
	\bibitem{AT92}
		G. S. Agarwal and K. Tara,
		Nonclassical character of states exhibiting no squeezing or sub-Poissonian statistics,
		\href{https://doi.org/10.1103/PhysRevA.46.485}{Phys. Rev. A \textbf{46}, 485 (1992)}.
	\bibitem{RV02}
		Th. Richter and W. Vogel,
		Nonclassicality of Quantum States: A Hierarchy of Observable Conditions,
		\href{http://dx.doi.org/10.1103/PhysRevLett.89.283601}{Phys. Rev. Lett. \textbf{89}, 283601 (2002)}.
	\bibitem{SVA16}
		J. Sperling, W. Vogel, and G. S. Agarwal,
		Operational definition of quantum correlations of light,
		\href{http://dx.doi.org/10.1103/PhysRevA.94.013833}{Phys. Rev. A \textbf{94}, 013833 (2016)}.
	\bibitem{S07}
		C. Silberhorn,
		Detecting quantum light,
		\href{http://dx.doi.org/10.1080/00107510701662538}{Contemp. Phys. \textbf{48}, 143 (2007)}.
	\bibitem{H09}
		R. H. Hadfield,
		Single-photon detectors for optical quantum information applications,
		\href{http://dx.doi.org/10.1038/nphoton.2009.230}{Nat. Photon. \textbf{3}, 696 (2009)}.
	\bibitem{LS99}
		A. Luis and L. L. S\'{a}nchez-Soto,
		Complete Characterization of Arbitrary Quantum Measurement Processes,
		\href{https://doi.org/10.1103/PhysRevLett.83.3573}{Phys. Rev. Lett. \textbf{83}, 3573 (1999)}.
	\bibitem{AMP04}
		G. M. D'Ariano, L. Maccone, and P. Lo Presti,
		Quantum Calibration of Measurement Instrumentation,
		\href{https://doi.org/10.1103/PhysRevLett.93.250407}{Phys. Rev. Lett. \textbf{93}, 250407 (2004)}.
	\bibitem{LKKFSL08}
		M. Lobino, D. Korystov, C. Kupchak, E. Figueroa, B. C. Sanders, and A. I. Lvovsky,
		Complete characterization of quantum-optical processes,
		\href{http://dx.doi.org/10.1126/science.1162086}{Science \textbf{322}, 563 (2008)}.
	\bibitem{LFCPSREPW09}
		J. S. Lundeen, A. Feito, H. Coldenstrodt-Ronge, K. L. Pregnell, C. Silberhorn, T. C. Ralph, J. Eisert, M. B. Plenio, and I. A. Walmsley,
		Tomography of quantum detectors,
		\href{http://dx.doi.org/10.1038/nphys1133}{Nat. Phys. \textbf{5}, 27 (2009)}.
	\bibitem{ZDCJEPW12}
		L. Zhang, A. Datta, H. B. Coldenstrodt-Ronge, X.-M. Jin, J. Eisert, M. B. Plenio, and I. A. Walmsley,
		Recursive quantum detector tomography,
		\href{http://dx.doi.org/10.1088/1367-2630/14/11/115005}{New J. Phys. \textbf{14}, 115005 (2012)}.
	\bibitem{BCDGMMPPP12}
		G. Brida, L. Ciavarella, I. P. Degiovanni, M. Genovese, A. Migdall, M. G. Mingolla, M. G. A. Paris, F. Piacentini, and S. V. Polyakov,
		Ancilla-Assisted Calibration of a Measuring Apparatus,
		\href{https://doi.org/10.1103/PhysRevLett.108.253601}{Phys. Rev. Lett. \textbf{108}, 253601 (2012)}.
	\bibitem{PHMH12}
		J. Pe\v{r}ina, O. Haderka, V. Mich\'{a}lek, and M. Hamar,
		Absolute detector calibration using twin beams,
		\href{https://doi.org/10.1364/OL.37.002475}{Opt. Lett. \textbf{37}, 2475 (2012)}.
	\bibitem{BKSSV17}
		M. Bohmann, R. Kruse, J. Sperling, C. Silberhorn, and W. Vogel,
		Direct calibration of click-counting detectors
		\href{https://doi.org/10.1103/PhysRevA.95.033806}{Phys. Rev. A \textbf{95}, 033806 (2017)}.
	\bibitem{LMN08}
		A. E. Lita, A. J. Miller, and S. W. Nam,
		Counting nearinfrared single-photons with 95\% efficiency,
		\href{https://doi.org/10.1364/OE.16.003032}{Opt. Express \textbf{16}, 3032 (2008)}.
	\bibitem{Getal11}
		T. Gerrits, \textit{et al}.,
		On-chip, photon-number-resolving, telecommunication-band detectors for scalable photonic information processing,
		\href{https://doi.org/10.1103/PhysRevA.84.060301}{Phys. Rev. A \textbf{84}, 060301(R) (2011)}.
	\bibitem{BCDGLMPRTP12}
		G. Brida, L. Ciavarella, I. P. Degiovanni, M. Genovese, L. Lolli, M. G. Mingolla, F. Piacentini, M. Rajteri, E. Taralli, and M. G. A. Paris,
		Quantum characterization of superconducting photon counters,
		\href{https://doi.org/10.1088/1367-2630/14/8/085001}{New J. Phys. \textbf{14}, 085001 (2012)}.
	\bibitem{RFZMGDFE12}
		J. J. Renema, G. Frucci, Z. Zhou, F. Mattioli, A. Gaggero, R. Leoni, M. J. A. de Dood, A. Fiore, and M. P. van Exter,
		Modified detector tomography technique applied to a superconducting multiphoton nanodetector,
		\href{http://dx.doi.org/10.1364/OE.20.002806}{Opt. Express \textbf{20}, 2806 (2012)}.
	\bibitem{ZCDPLJSPW12}
		L. Zhang, H. Coldenstrodt-Ronge, A. Datta, G. Puentes, J. S. Lundeen, X.-M. Jin, B. J. Smith, M. B. Plenio, and I. A. Walmsley,
		Mapping coherence in measurement via full quantum tomography of a hybrid optical detector,
		\href{https://doi.org/10.1038/nphoton.2012.107}{Nat. Photon. \textbf{6}, 364 (2012)}.
	\bibitem{LCGS10}
		K. Laiho, K. N. Cassemiro, D. Gross, and C. Silberhorn,
		Probing the Negative Wigner Function of a Pulsed Single Photon Point by Point,
		\href{https://doi.org/10.1103/PhysRevLett.105.253603}{Phys. Rev. Lett. \textbf{105}, 253603 (2010)}.
	\bibitem{BGGMPTPOP11}
		G. Brida, M. Genovese, M. Gramegna, A. Meda, F. Piacentini, P. Traina, E. Predazzi, S. Olivares, and M. G. A. Paris,
		Quantum state reconstruction using binary data from on/off photodetection,
		\href{http://dx.doi.org/10.1166/asl.2011.1204}{Adv. Sci. Lett. \textbf{4}, 1 (2011)}.
	\bibitem{LBFD08}
		E. Lantz, J.-L. Blanchet, L. Furfaro, and F. Devaux,
		Multi-imaging and Bayesian estimation for photon counting with EMCCDs,
		\href{http://dx.doi.org/10.1111/j.1365-2966.2008.13200.x}{Mon. Not. R. Astron. Soc. \textbf{386}, 2262 (2008)}.
	\bibitem{CWB14}
		R. Chrapkiewicz, W. Wasilewski, and K. Banaszek,
		High-fidelity spatially resolved multiphoton counting for quantum imaging applications,
		\href{http://dx.doi.org/10.1364/OL.39.005090}{Opt. Lett. \textbf{39}, 5090 (2014)}.
	\bibitem{ATDYRS15}
		M. J. Applegate, O. Thomas, J. F. Dynes, Z. L. Yuan, D. A. Ritchie, and A. J. Shields,
		Efficient and robust quantum random number generation by photon number detection,
		\href{http://dx.doi.org/10.1063/1.4928732}{Appl. Phys. Lett. \textbf{107}, 071106 (2015)}.
	\bibitem{WDSBY04}
		E. Waks, E. Diamanti, B. C. Sanders, S. D. Bartlett, and Y. Yamamoto,
		Direct Observation of Nonclassical Photon Statistics in Parametric Down-Conversion,
		\href{http://dx.doi.org/10.1103/PhysRevLett.92.113602}{Phys. Rev. Lett. \textbf{92}, 113602 (2004)}.
	\bibitem{HPHP05}
		O. Haderka, J. Pe\v{r}ina, Jr., M. Hamar, and J. Pe\v{r}ina,
		Direct measurement and reconstruction of nonclassical features of twin beams generated in spontaneous parametric down-conversion,
		\href{http://dx.doi.org/10.1103/PhysRevA.71.033815}{Phys. Rev. A \textbf{71}, 033815 (2005)}.
	\bibitem{FL13}
		R. Filip and L. Lachman,
		Hierarchy of feasible nonclassicality criteria for sources of photons,
		\href{https://doi.org/10.1103/PhysRevA.88.043827}{Phys. Rev. A \textbf{88}, 043827 (2013)}.
	\bibitem{APHAB16}
		I. I. Arkhipov, J. Pe\v{r}ina Jr., O. Haderka, A. Allevi, and M. Bondani,
		Entanglement and nonclassicality in four-mode Gaussian states generated via parametric down-conversion and frequency up-conversion,
		\href{http://dx.doi.org/10.1038/srep33802}{Sci. Rep. \textbf{6}, 33802 (2016)}.
	\bibitem{TKE15}
		S.-H. Tan, L. A. Krivitsky, and B.-G. Englert,
		Measuring quantum correlations using lossy photon-number-resolving detectors with saturation,
		\href{http://dx.doi.org/10.1080/09500340.2015.1076080}{J. Mod. Opt. \textbf{63}, 276 (2015)}.
	\bibitem{ALCS10}
		M. Avenhaus, K. Laiho, M. V. Chekhova, and C. Silberhorn,
		Accessing Higher Order Correlations in Quantum Optical States by Time Multiplexing,
		\href{http://dx.doi.org/10.1103/PhysRevLett.104.063602}{Phys. Rev. Lett. \textbf{104}, 063602 (2010)}.
	\bibitem{AOB12}
		A. Allevi, S. Olivares, and M. Bondani,
		Measuring high-order photon-number correlations in experiments with multimode pulsed quantum states,
		\href{http://dx.doi.org/10.1103/PhysRevA.85.063835}{Phys. Rev. A \textbf{85}, 063835 (2012)}.
	\bibitem{SBVHBAS15}
		J. Sperling, M. Bohmann, W. Vogel, G. Harder, B. Brecht, V. Ansari, and C. Silberhorn,
		Uncovering Quantum Correlations with Time-Multiplexed Click Detection,
		\href{http://dx.doi.org/10.1103/PhysRevLett.115.023601}{Phys. Rev. Lett. \textbf{115}, 023601 (2015)}.
	\bibitem{BDFL08}
		J.-L. Blanchet, F. Devaux, L. Furfaro, and E. Lantz,
		Measurement of Sub-Shot-Noise Correlations of Spatial Fluctuations in the Photon-Counting Regime,
		\href{http://dx.doi.org/10.1103/PhysRevLett.101.233604}{Phys. Rev. Lett. \textbf{101}, 233604 (2008)}.
	\bibitem{MMDL12}
		P.-A. Moreau, J. Mougin-Sisini, F. Devaux, and E. Lantz,
		Realization of the purely spatial Einstein-Podolsky-Rosen paradox in full-field images of spontaneous parametric down-conversion,
		\href{http://dx.doi.org/10.1103/PhysRevA.86.010101}{Phys. Rev. A \textbf{86}, 010101(R) (2012)}.
	\bibitem{CTFLMA16}
		V. Chille, N. Treps, C. Fabre, G. Leuchs, C. Marquardt, and A. Aiello,
		Detecting the spatial quantum uncertainty of bosonic systems,
		\href{https://doi.org/10.1088/1367-2630/18/9/093004}{New J. Phys. \textbf{18}, 093004 (2016)}.
	\bibitem{SBDBJDVW16}
		J. Sperling, T. J. Bartley, G. Donati, M. Barbieri, X.-M. Jin, A. Datta, W. Vogel, and I. A. Walmsley,
		Quantum Correlations from the Conditional Statistics of Incomplete Data,
		\href{http://dx.doi.org/10.1103/PhysRevLett.117.083601}{Phys. Rev. Lett. \textbf{117}, 083601 (2016)}.
	\bibitem{ZABGGBRP05}
		G. Zambra, A. Andreoni, M. Bondani, M. Gramegna, M. Genovese, G. Brida, A. Rossi, and M. G. A. Paris,
		Experimental Reconstruction of Photon Statistics without Photon Counting,
		\href{http://dx.doi.org/10.1103/PhysRevLett.95.063602}{Phys. Rev. Lett. \textbf{95}, 063602 (2005).}
	\bibitem{PADLA10}
		W. N. Plick, P. M. Anisimov, J. P. Dowling, H. Lee, and G. S. Agarwal,
		Parity detection in quantum optical metrology without number-resolving detectors,
		\href{http://dx.doi.org/10.1088/1367-2630/12/11/113025}{New J. Phys. \textbf{12}, 113025 (2010)}.
	\bibitem{KV16}
		B. K\"{u}hn and W. Vogel,
		Unbalanced Homodyne Correlation Measurements,
		\href{https://doi.org/10.1103/PhysRevLett.116.163603}{Phys. Rev. Lett. \textbf{116}, 163603 (2016)}.
	\bibitem{CKS14}
		M. Cooper, M. Karpinski, and B. J. Smith,
		Quantum state estimation with unknown measurements,
		\href{http://dx.doi.org/10.1038/ncomms5332}{Nat. Commun. \textbf{5}, 4332 (2014)}.
	\bibitem{AGSB16}
		M. Altorio, M. G. Genoni, F. Somma, and M. Barbieri,
		Metrology with Unknown Detectors,
		\href{https://doi.org/10.1103/PhysRevLett.116.100802}{Phys. Rev. Lett. \textbf{116}, 100802 (2016)}.
	\bibitem{PTKJ96}
		H. Paul, P. T\"{o}rm\"{a}, T. Kiss, and I. Jex,
		Photon Chopping: New Way to Measure the Quantum State of Light,
		\href{http://dx.doi.org/10.1103/PhysRevLett.76.2464}{Phys. Rev. Lett. \textbf{76}, 2464 (1996)}.
	\bibitem{KB01}
		P. Kok and S. L. Braunstein,
		Detection devices in entanglement-based optical state preparation,
		\href{http://dx.doi.org/10.1103/PhysRevA.63.033812}{Phys. Rev. A \textbf{63}, 033812 (2001)}.
	\bibitem{ASSBW03}
		D. Achilles, C. Silberhorn, C. \'{S}liwa, K. Banaszek, and I. A. Walmsley,
		Fiber-assisted detection with photon number resolution,
		\href{http://dx.doi.org/10.1364/OL.28.002387}{Opt. Lett. \textbf{28}, 2387 (2003)}.
	\bibitem{FJPF03}
		M. J. Fitch, B. C. Jacobs, T. B. Pittman, and J. D. Franson,
		Photon-number resolution using time-multiplexed single-photon detectors,
		\href{http://dx.doi.org/10.1103/PhysRevA.68.043814}{Phys. Rev. A \textbf{68}, 043814 (2003)}.
	\bibitem{RHHPH03}
		J. \v{R}eh\'{a}ček, Z. Hradil, O. Haderka, J. Pe\v{r}ina, Jr., and M. Hamar,
		Multiple-photon resolving fiber-loop detector,
		\href{http://dx.doi.org/10.1103/PhysRevA.67.061801}{Phys. Rev. A \textbf{67}, 061801(R) (2003)}.
	\bibitem{CDSM07}
		S. A. Castelletto, I. P. Degiovanni, V. Schettini, and A. L. Migdall,
		Reduced deadtime and higher rate photon-counting detection using a multiplexed detector array,
		\href{https://doi.org/10.1080/09500340600779579}{J. Mod. Opt. \textbf{54}, 337 (2007)}.
	\bibitem{SPDBCM07}
		V. Schettini, S.V. Polyakov, I.P. Degiovanni, G. Brida, S. Castelletto, and A.L. Migdall,
		Implementing a Multiplexed System of Detectors for Higher Photon Counting Rates,
		\href{https://doi.org/10.1109/JSTQE.2007.902846}{IEEE J. Sel. Top. Quantum Electron. \textbf{13}, 978 (2007)}.
	\bibitem{KK64}
		P. L. Kelley and W. H. Kleiner,
		Theory of Electromagnetic Field Measurement and Photoelectron Counting,
		\href{https://doi.org/10.1103/PhysRev.136.A316}{Phys. Rev. \textbf{136}, A316 (1964)}.
	\bibitem{I14}
		A. Ilyin,
		Generalized binomial distribution in photon statistics,
		\href{https://doi.org/10.1515/phys-2015-0005}{Open Phys. \textbf{13}, 41 (2014)}.
	\bibitem{PZA16}
		M. Pleinert, J. von Zanthier, and G. S. Agarwal,
		Quantum signatures of collective behavior of a coherently driven two atom system coupled to a single-mode of the electromagnetic field,
		\href{https://arxiv.org/abs/1608.00137}{arXiv:1608.00137 [quant-ph]}.
	\bibitem{MSB16}
		F. M. Miatto, A. Safari, and R. W. Boyd,
		Theory of multiplexed photon number discrimination,
		\href{https://arxiv.org/abs/1601.05831}{arXiv:1601.05831 [quant-ph]}.
	\bibitem{HBLNGS15}
		G. Harder, T. J. Bartley, A. E. Lita, S. W. Nam, T. Gerrits, and C. Silberhorn,
		Single-Mode Parametric-Down-Conversion States with 50 Photons as a Source for Mesoscopic Quantum Optics,
		\href{https://doi.org/10.1103/PhysRevLett.116.143601}{Phys. Rev. Lett. \textbf{116}, 143601 (2016)}.
	\bibitem{SVA12}
		J. Sperling, W. Vogel, and G. S. Agarwal,
		Sub-Binomial Light,
		\href{http://dx.doi.org/10.1103/PhysRevLett.109.093601}{Phys. Rev. Lett. \textbf{109}, 093601 (2012)}.
	\bibitem{BDJDBW13}
		T. J. Bartley, G. Donati, X.-M. Jin, A. Datta, M. Barbieri, and I. A. Walmsley,
		Direct Observation of Sub-Binomial Light,
		\href{http://dx.doi.org/10.1103/PhysRevLett.110.173602}{Phys. Rev. Lett. \textbf{110}, 173602 (2013)}.
	\bibitem{LFPR16}
		C. Lee, S. Ferrari, W. H. P. Pernice, and C. Rockstuhl,
		Sub-Poisson-Binomial Light,
		\href{https://doi.org/10.1103/PhysRevA.94.053844}{Phys. Rev. A \textbf{94}, 053844 (2016)}.
	\bibitem{MGHPGSWS15}
		T. Meany, M. Gr\"{a}fe, R. Heilmann, A. Perez-Leija, S. Gross, M. J. Steel, M. J. Withford, and A. Szameit,
		Laser written circuits for quantum photonics,
		\href{http://dx.doi.org/10.1002/lpor.201500061}{Laser Photon. Rev. \textbf{9}, 1863 (2015)}.
	\bibitem{HSPGHNVS16}
		R. Heilmann, J. Sperling, A. Perez-Leija, M. Gr\"afe, M. Heinrich, S. Nolte, W. Vogel, and A. Szameit,
		Harnessing click detectors for the genuine characterization of light states,
		\href{http://dx.doi.org/10.1038/srep19489}{Sci. Rep. \textbf{6}, 19489 (2016)}.

	\bibitem{AW70}
		G. S. Agarwal and E. Wolf,
		Calculus for Functions of Noncommuting Operators and General Phase-Space Methods in Quantum Mechanics. I. Mapping Theorems and Ordering of Functions of Noncommuting Operators,
		\href{https://doi.org/10.1103/PhysRevD.2.2161}{Phys. Rev. D \textbf{2}, 2161 (1970)};
		\textit{ibid.},
		II. Quantum Mechanics in Phase Space,
		\href{https://doi.org/10.1103/PhysRevD.2.2187}{Phys. Rev. D \textbf{2}, 2187 (1970)};
		\textit{ibid.},
		III. A Generalized Wick Theorem and Multitime Mapping,
		\href{https://doi.org/10.1103/PhysRevD.2.2206}{Phys. Rev. D \textbf{2}, 2206 (1970)}.
	\bibitem{VW06}
		See Ch. 8 in W. Vogel and D.-G. Welsch,
		\textit{Quantum Optics}
		(Wiley-VCH, Weinheim, 2006).
	\bibitem{LSV15}
		T. Lipfert, J. Sperling, and W. Vogel,
		Homodyne detection with on-off detector systems,
		\href{http://dx.doi.org/10.1103/PhysRevA.92.053835}{Phys. Rev. A \textbf{92}, 053835 (2015)}.
	\bibitem{SVA13}
		J. Sperling, W. Vogel, and G. S. Agarwal,
		Correlation measurements with on-off detectors,
		\href{http://dx.doi.org/10.1103/PhysRevA.88.043821}{Phys. Rev. A \textbf{88}, 043821 (2013)}.
	\bibitem{BKSSV17atm}
		M. Bohmann, R. Kruse, J. Sperling, C. Silberhorn, and W. Vogel,
		Probing free-space quantum channels with in-lab experiments,
		\href{https://arxiv.org/abs/1702.04127}{arXiv:1702.04127 [quant-ph]}.
	\bibitem{ZM90}
		X. T. Zou and L. Mandel,
		Photon-antibunching and sub-Poissonian photon statistics,
		\href{https://doi.org/10.1103/PhysRevA.41.475}{Phys. Rev. A \textbf{41}, 475 (1990)}.
	\bibitem{ECMS11}
		A. Eckstein, A. Christ, P. J. Mosley, and C. Silberhorn,
		Highly Efficient Single-Pass Source of Pulsed Single-Mode Twin Beams of Light,
		\href{https://doi.org/10.1103/PhysRevLett.106.013603}{Phys. Rev. Lett. \textbf{106}, 013603 (2011)}.
	\bibitem{LCS09}
		K. Laiho, K. N. Cassemiro, and Ch. Silberhorn,
		Producing high fidelity single photons with optimal brightness via waveguided parametric down-conversion,
		\href{https://doi.org/10.1364/OE.17.022823}{Opt. Express \textbf{17}, 22823 (2009)}.
	\bibitem{KHQBSS13}
		S. Krapick, H. Herrmann, V. Quiring, B. Brecht, H. Suche and Ch. Silberhorn,
		An efficient integrated two-color source for heralded single photons,
		\href{http://dx.doi.org/10.1088/1367-2630/15/3/033010}{New J. Phys. \textbf{15}, 033010 (2013)}.
	\bibitem{MLCVGN11}
		A. J. Miller, A. E. Lita, B. Calkins, I. Vayshenker, S. M. Gruber, and S. W. Nam,
		Compact cryogenic self-aligning fiber-to-detector coupling with losses below one percent,
		\href{https://doi.org/10.1364/OE.19.009102}{Opt. Express \textbf{19}, 9102 (2011)}.
	\bibitem{I95}
		K. D. Irwin,
		An application of electrothermal feedback for high resolution cryogenic particle detection,
		\href{http://dx.doi.org/10.1063/1.113674}{Appl. Phys. Lett. \textbf{66}, 1998 (1995)}.
	\bibitem{HMGHLNNDKW15}
		P. C. Humphreys, B. J. Metcalf, T. Gerrits, T. Hiemstra, A. E. Lita, J. Nunn, S. W. Nam, A. Datta, W. S. Kolthammer, and I. A. Walmsley,
		Tomography of photon-number resolving continuous-output detectors,
		\href{http://dx.doi.org/10.1088/1367-2630/17/10/103044}{New J. Phys. \textbf{17}, 103044 (2015)}.
	\bibitem{SVA14}
		J. Sperling, W. Vogel, and G. S. Agarwal,
		Quantum state engineering by click counting,
		\href{http://dx.doi.org/10.1103/PhysRevA.89.043829}{Phys. Rev. A \textbf{89}, 043829 (2014)}.

\end{thebibliography}
\end{document}